\documentclass[%
 reprint,
 superscriptaddress,
 amsmath,amssymb,
 aps,
 prb,
 longbibliography,
]{revtex4-2}

\usepackage{graphicx}
\usepackage{amsmath}
\usepackage{amssymb}
\usepackage{dcolumn}
\usepackage{bm}
\usepackage{physics}
\usepackage[colorlinks=true,citecolor=RoyalBlue,linkcolor=RoyalBlue,urlcolor=RoyalBlue]{hyperref}
\usepackage{url}
\usepackage[dvipsnames]{xcolor}
\usepackage{mathrsfs}
\usepackage{mathtools}
\usepackage{braket}
\usepackage[T1]{fontenc} 
\usepackage[normalem]{ulem}

\newcommand{\mi}{\mathrm{i}}
\renewcommand{\vec}[1]{\textbf{\textit{#1}}}

\newcommand{\muB}[0]{\mu_\text B}

\newcommand{\edit}[2]{{\ifmmode\text{\sout{\ensuremath{#1}}}\else\sout{#1}\fi}{\textcolor{Red}{#2}}}

\begin{document}

\title{Four-state discrimination for a pair of spin qubits via gate reflectometry}

\author{Aritra Sen}
\affiliation{%
 Department of Theoretical Physics, Institute of Physics, Budapest University of Technology and Economics, M\H{u}egyetem rkp. 3., H-1111 Budapest, Hungary
}

\author{Andr\'as P\'alyi}
\affiliation{%
 Department of Theoretical Physics, Institute of Physics, Budapest University of Technology and Economics, M\H{u}egyetem rkp. 3., H-1111 Budapest, Hungary
}%
\affiliation{HUN-REN-BME-BCE Quantum Technology Research Group, Budapest University of Technology and Economics, M\H{u}egyetem rkp. 3., H-1111 Budapest, Hungary}

\date{\today}

\begin{abstract}
Single-electron spin qubits defined in quantum dots are used as building blocks of a semiconductor-based quantum computer.
Readout in a scaled-up version of such a quantum computer is expected to rely on the Pauli Spin Blockade (PSB) mechanism. 
A desired functionality of PSB readout is that it reveals two bits of information on the two spin qubits that are involved in the process, such that the four computational basis states can be discriminated.
In this work, we propose and quantitatively analyze an experimental procedure, based on gate reflectometry, which enables this four-state discrimination in a single measurement.
We provide an intuitive recipe to maximize the contrast between the quantum capacitances of the four basis states. 
Focusing on silicon double quantum dots equipped with a micromagnet, we quantify how amplifier noise and phonon-mediated relaxation influence readout fidelity.
Our results highlight a realistic opportunity to mitigate the overhead of readout ancilla qubits in a spin-based quantum computer.
\end{abstract}

\maketitle

\tableofcontents

\section{Introduction}

Semiconductor quantum dot spin qubits have emerged as an important platform for quantum information processing, owing to their long coherence times, compatibility with industrial fabrication, and potential for dense integration \cite{Loss1998,HansonReview,Stano2022,Zwanenburg2013SiQuantElec,Burkard2023review,ChatterjeeReview}.
The early proposal \cite{Loss1998} identified the spin degree of freedom of a confined electron as a natural two-level system for quantum computation, and subsequent experiments demonstrated readout \cite{Elzerman2004} and coherent control \cite{Petta2005,KoppensNature2006} of single spins in gate-defined quantum dots. 
Over the past decade, rapid progress has been achieved in high-fidelity single- \cite{Yoneda2018,YiHsienWuArxiv2025} and two-qubit gates \cite{HuangNature2019,Xue2022,Steinacker2025}, the implementation of small multi-qubit processors \cite{Philips2022,FernndezdeFuentes2026,HendrickxNature2021,XinZhang2025}, and elementary quantum simulation \cite{DehollainNagaoka,vanDiepenSimulation,HsiaoExciton} and error-correction primitives \cite{vanRiggelenQEC2022,TakedaQEC2022,UndsethArxiv2026}. 
In particular, silicon-based devices benefit from the availability of isotopically enriched material, which strongly suppresses hyperfine-induced decoherence \cite{VeldhorstAddressable,VeldhorstTwoQubitLogic,Muhonen30sec,DehollainBell,Yoneda2021,Yoneda2018,Tyryshkin2012,ItohReview2014,Mills2022}, and from their compatibility with established semiconductor technology \cite{Steinacker2025,Zwerver2022,Neyens2024,George2025,MadzikNature2025,KotekarPatilPSS2017,Maurand2016cmos,Urdampilleta2019CMOSGateRf,Crippa2019reflectometrySi,Bartee2025}. 
Complemented with an increasing activity in architecture design \cite{Boter2022,Pataki2025,vandersypen2017Scalling,Kunne2024SpinBus,Patomki2024SiPipeline}, these advances position semiconductor spin qubits as a promising route toward large-scale quantum computing and quantum simulation architectures.

A central ingredient of any scalable architecture is high-fidelity qubit readout. In a double quantum dot (DQD) with two spin qubits, readout is commonly realized via the Pauli spin blockade (PSB) mechanism \cite{OnoScience2002,KoppensNature2006}, which maps spin information onto charge configurations \cite{Petta2005} upon tuning the DQD from the (1,1) to the (0,2) charge configuration. 
In its conventional implementation, the spin-dependent tunneling between the (1,1) and (0,2) charge states results in a blockade for triplet configurations, while singlet states hybridize with (0,2) and can tunnel.
This spin-to-charge conversion is typically detected by a proximal charge sensor, such as a quantum point contact \cite{Elzerman2004,SchleserAPL2004,Petta2005,Reilly2007RfQPC}, a single-electron transistor \cite{Barthel2010,BuehlerAPL2005,Jang2020RfSET,Nurizzo2023,Huang2024Nature1KSiMOS}, or a single-electron box \cite{Ciriano-Tejel2021,Laine2025dispersiveHMM}.

Besides charge sensing, another way to identify different two-electron energy eigenstates in a DQD is to measure the quantum capacitance, i.e., the amount of charge redistribution in the DQD in response to a small change of a plunger gate voltage.
The quantum capacitance can be measured, e.g., through the reflectance of a tank circuit in which the DQD is connected through its plunger gate \cite{Vigneau2023}.  
This method is often referred to as \emph{gate reflectometry}.

Gate reflectometry provides an alternative way of spin qubit readout.
First, the spin information is
 converted into quantum capacitance by tuning the DQD from (1,1) into the vicinity of the (1,1)-(0,2) charge transition point; then, the quantum capacitance of the state is probed by dispersive gate-based reflectometry \cite{CollessPRL2013,West2019SingleShotRF,Crippa2019reflectometrySi,Connors2020GateRfRapid,Vigneau2023,Mizuta2017,Esterli2019,Peri2024Unified,Urdampilleta2019CMOSGateRf,Russell2023GateRfSiHole,Zheng2019rapidreadout}. 
Towards scaling, gate reflectometry has the advantage that it avoids the footprint and wiring overhead of a nearby charge sensor.
Here, we use the term `PSB readout' in a broad sense, including charge sensing or capacitance measurement, as well as cases where spin-orbit interaction \cite{PfundPRL2007,Danon2009psbStrongSOC,ScherublCommsPhys2019,Russell2023GateRfSiHole,Sen2023,Lundberg2024LiftedPSB,vonHorstig2025}, hyperfine interaction \cite{Petta2005,KoppensNature2006,Jouravlev2006Transport}, or inhomogeneous magnetic fields \cite{PioroLadriereNatPhys2008,Unseld2025} break the spin selection rules.

PSB readout has three major modalities: 
(i) ancilla mode, where one of the two spins has a known state, and is used as a reference or ancilla for reading out the other spin; 
(ii) singlet-triplet mode, where the (1,1) singlet state is discriminated from the (1,1) triplet states; and
(iii) parity mode, where the parallel-spin states (i.e., the polarized triplets) are distinguished from the antiparallel-spin states (i.e., the singlet and the unpolarized triplet).

In these modalities, PSB readout results in a single bit of information about the two spins.
Is it possible to improve on this, i.e., to gain two bits, and hence discriminate all four spin states?
A recent experiment \cite{Nurizzo2023} has demonstrated this functionality in a GaAs device by performing a sequence of three manipulation and readout steps, using optimized readout positions in the detuning-tunneling parameter space, careful control pulses to navigate between those positions, and charge sensing with a single electron transistor.
Another experimental work \cite{Laine2025dispersiveHMM} achieved three-state discrimination in a Si device, using a single electron box charge sensor, and implementing state discrimination using a hidden Markov model.

In this work, we propose, model, and quantitatively analyze a method to simultaneously read out two single-electron spin qubits (i.e, Loss-DiVincenzo \cite{Loss1998} spin qubits) in a double quantum dot.
We propose to achieve this by discriminating the four spin states in PSB readout, by  measuring the quantum capacitance of the DQD with gate reflectometry in a single shot.
The method we describe here relies on the presence of level anticrossings and spin and charge hybridization in the detuning- and tunneling-dependent energy spectrum of the two-electron spin states, which in turn are caused by interdot tunneling and spin-mixing effects such as spin-orbit interaction or an inhomogeneous magnetic field.
We focus on a Si-based two-electron DQD in the presence of a micromagnet, the latter enabling electrically driven spin control \cite{PioroLadriereNatPhys2008}, and also opening the opportunity to four-state discrimination (4SD).
We describe a way to maximize the contrast between the quantum capacitances of the four spin states.
We also evaluate the contributions of phonon-mediated relaxation and amplifier noise to the readout error, and illustrate how the readout position can be optimized in the presence of those imperfections.

The rest of this paper is structured as follows. 
In Sec.~\ref{sec:model} we specify our model Hamiltonian for the two-electron DQD in presence of inhomogeneous magnetic field. 
In Sec.~\ref{sec:four-state-discrimination} we describe the idea of four-state discrimination of a pair spin qubits using quantum capacitance measurement, and describe the figures of merits in terms of the capacitance contrast and the readout error. 
In Sec.~\ref{sec:relaxation} we describe relaxation effects and how it influences readout error and the optimal readout position.
Finally, we provide a discussion and outlook in Sec.~\ref{sec:discussion}, and conclude in Sec.~\ref{sec:conclusions}.
The Appendix contains details of our calculations.

\begin{figure}
\centering
\includegraphics[width=1\linewidth]{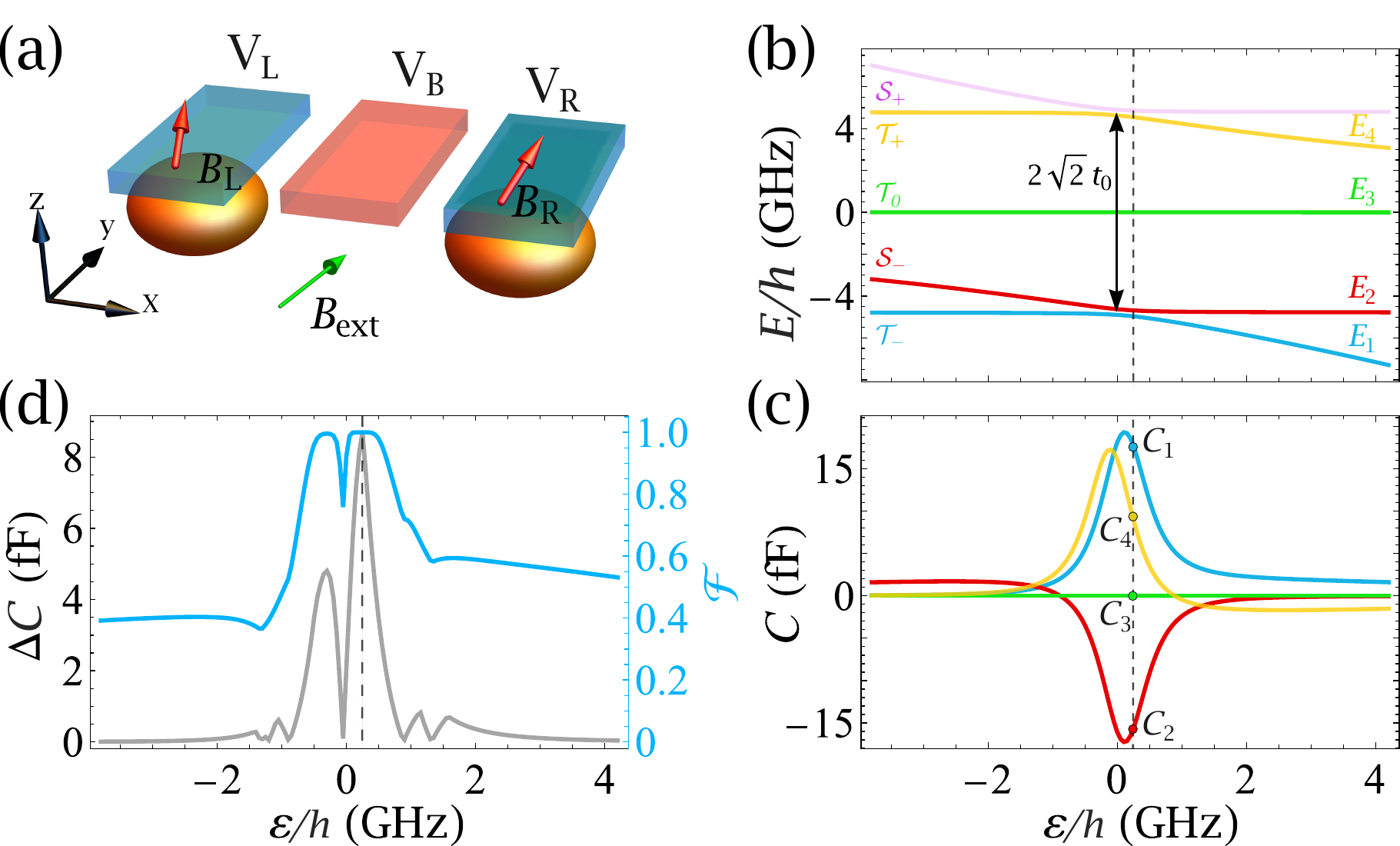}
\caption{\textbf{Four-state discrimination with a single quantum capacitance measurement.} 
(a) Schematics of the double quantum dot (DQD) in the presence of an inhomogeneous magnetic field.
Energy spectrum (b) of the DQD, and quantum capacitances (c) of the energy eigenstates, as functions of detuning, for $t_0/h =  3.35$ GHz. 
Internal magnetic fields $\vec B_L$, $\vec B_R$ in the two dots are different, and emerge as combinations of the homogeneous external magnetic field $B_\mathrm{ext}$ and the inhomogeneous micromagnet field. 
Magnetic-field parameters, related to $\vec B_L$ and $\vec B_R$ via Eqs.~\eqref{eq:Bs}, \eqref{eq:Ba}, \eqref{eq:Bsa}:
$B_s = 171.8$ mT,
$B_{a\parallel} = -2.5$ mT,
and
$B_{a\perp} = 9.8$ mT.
Parameters used for (b,c,d) are in the range $t_0 \approx g \muB B_s / \sqrt{2}$, cf.~Eq.~\eqref{eq:alignment_condition}.
In (b), labels on the left identify the spin character of the states; 
$\mathcal{T}_-$, $\mathcal{T}_+$:
polarized (1,1) triplets; 
$\mathcal{S}_-$: bonding singlet;
$\mathcal{S}_+$: antibonding singlet;
$\mathcal{T}_0$: unpolarized (1,1) triplet.
(d) Quantum capacitance contrast $\Delta C$ (gray),  and four-state assignment fidelity 
$\mathcal{F}$
(blue), the latter computed with amplifier noise strength of $\sigma_C = 1$ fF. 
Gray dashed line in panels b, c, d indicate the maximum point of the capacitance contrast $\Delta C(\varepsilon)$.
Quantum capacitance of the pink state is not shown, since that state is not adiabatically connected to the four computational basis states, and has higher energy than those four states; hence we do not expect it to be populated.
}
\label{fig:setup}
\end{figure}

\section{Double quantum dot Hamiltonian with two electrons}
\label{sec:model}

We focus on 4SD by a quantum capacitance measurement, in a two-electron DQD equipped with a micromagnet.
The DQD with the micromagnet field is depicted in Fig.~\ref{fig:setup}(a).
The micromagnet enables electrically driven spin resonance via gate-voltage microwave bursts. 
As we argue here, the presence of the micromagnet also opens an opportunity for 4SD.

The Hamiltonian of the DQD reads as follows:
\begin{subequations}  
\label{eq:Hamiltonian}
    \begin{align}
        & H =   H_\text{Coulomb} +  H_\text{on-site} +  H_\text{tunneling} +  H_\text{Zeeman},\label{eq:twoelectronhamiltonian}  \\
        & H_\text{Coulomb} = U( n_{\text L \uparrow}  n_{\text L \downarrow} +  n_{\text R \uparrow} n_{\text R \downarrow}),\label{eq:hcoulomb} \\ 
        & H_\text{on-site} = \sum_{j=\text{L,R}} \varepsilon_j  n_j,\label{eq:honsite} \\ 
        & H_\text{tunneling} = t_0\sum_{s=\uparrow,\downarrow} c^\dagger_{\text L, s} c_{\text R, s} + h.c.,\label{eq:htunneling} \\
        & H_\text{Zeeman} = \frac 1 2 g \muB 
        \left(
        \boldsymbol{B}_\mathrm{L} \cdot\boldsymbol{\sigma}_\mathrm{L}+
        \boldsymbol{B}_\mathrm{R} \cdot \boldsymbol{\sigma}_\mathrm{R}
        \right),
    \end{align} 
\end{subequations}
Here, 
$j\in\{\text{L,R}\}$ is the dot label,
$s\in\{{\uparrow,\downarrow}\}$ labels the two spin basis states, 
$c_{j,s}$ is the electron annihilation operator of dot $j$ and spin $s$,
$n_{j,s} = c^\dagger_{j,s}c_{j,s}$ is the number of electrons on dot $j$ with spin $s$, and $n_j=\sum_s n_{j,s}$. 
The on-site Coulomb repulsion energy is denoted by $U$.
The on-site energies of the dots are denoted by $\varepsilon_j$, and we define $\tilde \varepsilon = \varepsilon_\mathrm{L} -\varepsilon_\mathrm{R}$ and $\overline\varepsilon = (\varepsilon_\mathrm{L} + \varepsilon_\mathrm{R})/2$. 
Later, we will use the \emph{detuning} defined as $\varepsilon = \tilde \varepsilon - U$, i.e., $\varepsilon$ is the detuning from the (1,1)-(0,2) charge transition point. 
To make our plots, we will set  $\overline\varepsilon=0$. The spin operators $\boldsymbol\sigma_j$ are Pauli spin operators of the two dots; e.g., $\sigma_{\mathrm{L},z} = 
 n_{\mathrm{L},\uparrow} -
 n_{\mathrm{L},\downarrow}
$.

For simplicity, we assume a trivial structure for the gate-voltage lever arms. 
In particular, we consider the quantum capacitance of the two-electron states with respect to the plunger gate of dot $R$, assuming a simple relation $\varepsilon_\mathrm{R} = -|e| V_R$, i.e., assuming unit lever arm between plunger $R$ and dot $R$, and neglecting the cross-coupling between $V_R$ and other DQD parameters. 
Generalization of our considerations to a more realistic lever arm structure is straightforward \cite{Mizuta2017,Vigneau2023}.

For convenience, we work in a symmetrized magnetic reference frame \cite{Jouravlev2006Transport,TaylorPRB2007,DanonPRB2013}. That is, we align the magnetic $\tilde{z}$ axis with the average magnetic field of the two dots, 
\begin{equation}
\label{eq:Bs}
\boldsymbol{B}_{s} = (\boldsymbol{B}_L + \boldsymbol{B}_R)/2,
\end{equation}
and define the magnetic $\tilde{x}$ axis such that the difference 
\begin{equation}
\label{eq:Ba}
\boldsymbol{B}_a = (\boldsymbol{B}_L - \boldsymbol{B}_R)/2
\end{equation}
of the local magnetic fields lies in the $\tilde{x}>0$ half-plane of the 
$\tilde{x}$-$\tilde{z}$ plane.
This implies that the symmetrized and antisymmetrized magnetic field takes the following form in the $\tilde{x}$-$\tilde{y}$-$\tilde{z}$  magnetic reference frame: 
\begin{equation}
\label{eq:Bsa}
\vec{B}_s = (0,0,B_s),
\mbox{ and }
\vec{B}_a = (B_{a\perp},0,B_{a\parallel}),
\end{equation}
with $B_{a\perp} > 0$.
For the rest of the paper, we use example parameter sets where $B_{a\perp}$ and $B_{a\parallel}$ take values of a few milliteslas, whereas $B_s \sim 170$ mT, estimated from data on a micromagnet setup with an in-plane external field of $B_\mathrm{ext} = 70$ mT \cite{Unseld2025,unseld2025Zenodo}.
With this parameter set, the Larmor frequency difference of the two spin qubits is sufficiently large to enable their selective addressing  \cite{Unseld2025} with microwaves.

Throughout the paper, we focus on the vicinity of the (1,1)-(0,2) charge transition and the range of moderate tunneling energies, that is, $|\varepsilon|, |t_0| \ll U$.
Hence, we disregard the (2,0) singlet state, and will use the $5 \times 5$ matrix representation of the Hamiltonian of Eq.~\eqref{eq:Hamiltonian}.
In the singlet-triplet basis of the (0,2) singlet $\ket{\mathcal{S}}$, and the (1,1) states
$\ket{\mathcal S_0}$, 
$\ket{\mathcal T_0}$,
$\ket{\mathcal T_+}$,
$\ket{\mathcal T_-}$, this matrix representation reads:
\begin{equation}\label{eq:H5}
    H = \mqty(-\varepsilon & \sqrt 2 t_0 & 0 & 0 & 0\\
                \sqrt 2 t_0 & 0 & g\muB B_{a\parallel} & -\frac{g\muB B_{a\perp}}{\sqrt 2}  & \frac{g\muB B_{a\perp}}{\sqrt 2} \\
                0 & g\muB B_{a\parallel} & 0 & 0 & 0\\
                0 & -\frac{g\muB B_{a\perp}}{\sqrt 2} & 0 & g\muB B_s & 0\\
                0 & \frac{g\muB B_{a\perp}}{\sqrt 2} & 0 & 0 & -g\muB B_s).
\end{equation}

\section{Four-state discrimination with quantum capacitance measurement}
\label{sec:four-state-discrimination}

In a spin-based quantum computer, spin readout can be carried out in various ways. Here, we focus on a situation when two spin qubits are located in two neighboring, but separated, quantum dots, the state of this two-qubit system is unknown to us, and we wish to read them out in the computational basis.
This four-state discrimination task can be done via a quantum capacitance measurement, as we describe in this section.

\subsection{Capacitance contrast as a proxy for the quality of four-state discrimination}
\label{subsec:adiabatic-capacitance-contrast}

Before starting the readout process, we assume that the tunnel coupling between the dots is switched off, $t_0 = 0$, and the detuning parameter is tuned to the center of the (1,1) charge configuration, that is, $\varepsilon_L = \varepsilon_R = 0$ and hence $\varepsilon = -U$.
In this position, single-qubit gates can be performed by microwave gate-voltage bursts, selectively addressing the two qubits due to their Larmor-frequency difference caused by the micromagnet.
Two-qubit gates can also be performed, by activating the tunnel coupling between the two dots.
We call this the idle position.
Our target is to read out \emph{both qubits} in a single shot.
This readout is done in a different parameter point: detuning is tuned to the vicinity of the (1,1)-(0,2) charge transition, $\varepsilon \approx 0$, and the interdot tunnel coupling $t_0$ is switched on.
We call this the readout position.
We plot the energy eigenvalues of the Hamiltonian \eqref{eq:twoelectronhamiltonian} as a function of detuning $\varepsilon$ in a range of possible readout positions, in Fig.~\ref{fig:setup}(b), for a specific set of model parameters (see caption).
Simultaneous readout of the two qubits is achieved by a measurement that discriminates the four lowest-energy states in Fig.~\ref{fig:setup}(b).

In this work, we assume that when moving from the idle position to the readout position, the detuning and the tunnel coupling are tuned in a way that their change leads to a perfect state transfer.
That is, each computational basis state of the idle position evolves into the corresponding energy eigenstate in the readout position.
This is enabled by the absence of level crossings in the parameter-dependent energy spectrum, which, in turn, is assured by the misaligned local magnetic fields in the two dots. 
Note, however, that optimizing such a high-fidelity state transfer protocol is in general a challenging problem \cite{Fehse2023StateTransfer,KatiraeeFar2025Optimization,Ventura-Meinersen2025Optimization}.

The simultaneous readout of the two qubits can realized, if the four lowest-energy eigenstates at the readout position have significantly different quantum capacitances.
The quantum capacitance of a state can be measured by gate reflectometry, e.g., by probing the reflectance of a resonant circuit (e.g., a lumped-element tank circuit, or a superconducting resonator) that is connected to a plunger gate of one of the dots \cite{West2019SingleShotRF,Zheng2019rapidreadout,Crippa2019reflectometrySi,Liu2021,Connors2022,Urdampilleta2019CMOSGateRf,Esterli2019,Mizuta2017,Peri2024Unified,Vahid2020,Vigneau2023}.

Fig.~\ref{fig:setup}(c) shows the quantum capacitances of all four relevant states as the functions of the detuning parameter in a range of readout positions, for a specific parameter set (see caption).
We label the lowest 4 energy eigenstates, corresponding to the two-qubit computational basis, with $j \in \{1,2,3,4\}$, in ascending order of their energies, that is, $E_1 < E_2 < E_3 < E_4$.
The quantum capacitance $C_j$ of energy eigenstate $\ket{\psi_j}$, sometimes also referred to as the charge susceptibility, describes the `floppyness' of the charge, i.e., the change of charge with respect to a change of the gate voltage.
For our case (recall that we assume a trivial lever arm structure), this quantum capacitance is expressed as:
\begin{equation}\label{eq:quantumcapacitance_charge_derivative}
C_j = \frac{d Q_{R,j}}{d V_R},
\end{equation}
where $Q_{R,j} = |e| \braket{\psi_j | n_R | \psi_j }$ is the gate charge on plunger R.
Together with $\varepsilon_\mathrm{R} = -|e| V_R$ and 
$\varepsilon = \varepsilon_\mathrm{L}- \varepsilon_\mathrm{R}- U$, these imply the following simple formula:
\begin{equation}
\label{eq:quantumcapacitance}
C_j = e^2 \frac{d \braket{\psi_j | n_{\mathrm{R}} | \psi_j}}{d \varepsilon}.
\end{equation}
Fig.~\ref{fig:setup}(c) demonstrates that the readout position can be chosen such that the quantum capacitances of all four states are different at the readout position. 
Such a detuning value is indicated by the dashed line in Fig.~\ref{fig:setup}(c) (and also shown in panels b and d).

Instrumental noise adds uncertainty to the measured value of the quantum capacitance. 
Therefore, for high-fidelity state assignment, it is beneficial to have the four capacitance values as far from each other as possible. 
This implies that a reasonable proxy for the quality of four-state discrimination is the minimum of capacitance differences, which we call the \emph{capacitance contrast}:
\begin{equation}
    \Delta C = 
    \min_{i,j \in \{1,2,3,4\},\,  i<j} \left| C_j - C_i \right|.
\end{equation}
The dependence of the capacitance contrast on the detuning value for our example parameter set is shown as the gray line in Fig.~\ref{fig:setup}(d).
As Fig.~\ref{fig:setup}(d) reveals, the dashed lines in Fig.~\ref{fig:setup}(a,b,c) show the detuning value where four-state discrimination has optimal (maximal) capacitance contrast for this example parameter set.

\subsection{Assignment fidelity of four-state discrimination in the presence of amplifier noise}
\label{sec:amplifier-noise}

Upon readout, the quantum capacitance of a state cannot be inferred with absolute precision, hence the four-state discrimination procedure will also be noisy. 
Uncertainty of the measured quantum capacitance is caused, e.g., by amplifier noise, which is often described as a Gaussian broadening of the reflectance peak, whose center is determined by the quantum capacitance \cite{Vahid2020,Zheng2019rapidreadout,Barthel2009SingleShot,Vigneau2023}.

\begin{figure}[t]
    \centering
\includegraphics[width=1\linewidth]{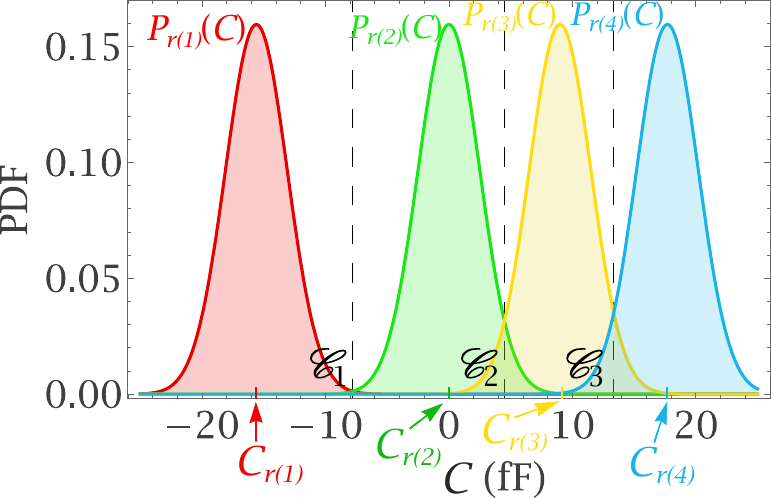}
    \caption{\textbf{Four-state discrimination and readout error in the presence of amplifier noise.} 
    Colored curve $P_{r(j)}$ is the probability density function of the quantum capacitance value inferred from a reflectance measurement performed on state $j$, in the presence of amplifier noise.
    Vertical dashed lines are separators used for maximum-likelihood inference, separating the intervals defined in Eq.~\eqref{eq:intervals}.
    Means of the probability density functions correspond to the capacitances in Fig.~\ref{fig:setup}(c) along the dashed vertical line:  $C_{r(j)}=\{-15.6, 0, 9.0, 17.7\}$ fF. Amplifier noise strength: $\sigma_C = 2.5$ fF.
    The readout error is determined by the overlaps between the neighboring Gaussians.}
    \label{fig:inferenceexplained}
\end{figure}

In this subsection, we quantify the effect of this uncertainty on the quality of four-state discrimination.
We assume that that if the computational basis state $j$ is being measured by reflectometry, then the inferred value of the capacitance is drawn from the Gaussian distribution
\begin{equation}
\label{eq:gaussian}
P_j(C) = \frac{1}{\sqrt{2\pi} \sigma_C}
\exp\left({-\frac{(C-C_j)^2}{2\sigma_C^2}}\right).    
\end{equation}
Note that the broadening, or \emph{amplifier noise strength}, $\sigma_C$ is assumed to be the same for all four states.
These Gaussian distributions are shown in Fig.~\ref{fig:inferenceexplained}, for noise strength $\sigma_C = 2.5$ fF, for parameters corresponding to the dashed vertical line in Fig.~\ref{fig:setup}(c).

In the presence of the above-mentioned uncertainty of inferring the quantum capacitance from a reflectometry measurement, it is natural to use the maximum likelihood principle to define the inference rule for 4SD.
Informally, this rule is that if we infer the quantum capacitance value of $C$, then our readout procedure outputs the state label $j$ whose likelihood $P_j(C)$ is maximal.

To formalize the maximum likelihood principle, we introduce the following simple ranking function $r$. It is a permutation of the state labels, $r:\{1,2,3,4\} \to \{1,2,3,4\}$, which ranks the capacitances in ascending order: $C_{r(1)} < C_{r(2)} < C_{r(3)} < C_{r(4)}$. 
For example, the ranking function corresponding to the parameter set of the dashed line in Fig.~\ref{fig:setup}(c) is $r(1) = 2$, 
$r(2) = 3$, 
$r(3) = 4$, 
$r(4) = 1$, implying that indeed, it holds that $C_{r(1)} < C_{r(2)} < C_{r(3)} < C_{r(4)}$, since it holds that 
$C_2 < C_3 < C_4 < C_1$.

With this notation, the inference rule can be reformulated as follows:
\begin{equation}
\label{eq:inferencerule}
\mathcal{I}: C \mapsto r(j), \mbox{ if }C \in J_{r(j)},
\end{equation}
where
\begin{subequations}
\label{eq:intervals}
\begin{eqnarray}
J_{r(1)} &=& ]-\infty,\mathscr{C}_1],
\\
J_{r(2)} &=& ]
\mathscr{C}_1,
\mathscr{C}_2,
],
\\
J_{r(3)} &=& ]
\mathscr{C}_2,
\mathscr{C}_3
],
\\
J_{r(4)} &=& ] \mathscr{C}_3, \infty [,
\end{eqnarray}
\end{subequations}
and $\mathscr{C}_i=(C_{r(i)} + C_{r(i+1)})/2$.
The threshold values $\mathscr{C}_1$, $\mathscr{C}_2$, and $\mathscr{C}_3$, 
separating the four intervals, are shown as the vertical dashed lines in Fig.~\ref{fig:inferenceexplained}.

The probability of correct inference of the state $j$ is therefore:
\begin{equation}\label{eq:fidelity_each_state}
\mathcal{F}_{j} = \int_{J_{r(j)}} dC P_{j}(C).
\end{equation}
To quantify the quality of 4SD, we will use the \textit{assignment fidelity}, or \emph{fidelity} for short, which we define as the probability of correct inference, averaged over the four states:
\begin{equation}
\label{eq:fidelity}
\mathcal{F} = \frac 1 4 \sum_{j=1}^4
    \mathcal{F}_{j}.
\end{equation}
Alternatively, the quality of 4SD can also be expressed by $1-\mathcal{F}$, which we call the \emph{assignment infidelity}, or simply \emph{readout error}. 
Combining Eqs.~\eqref{eq:gaussian},
\eqref{eq:inferencerule}, 
\eqref{eq:intervals},
and \eqref{eq:fidelity_each_state} with the definition of Eq.~\eqref{eq:fidelity}, the assignment fidelity can also be expressed as: 
\begin{eqnarray}\label{eq:average_assignment_fidelity}
    \nonumber
    \mathcal{F} &=& 
    \frac{1}{4}\left[\Phi(\mathscr{C}_{1}- C_{r(1)}) \right.
    \\
    \nonumber
    &+& \left(
    \Phi(\mathscr{C}_{2}-C_{r(2)})
    -\Phi(\mathscr{C}_{1}-C_{r(2)})
    \right)
    \\
    \nonumber
    &+&  \left(
    \Phi(\mathscr{C}_{3}-C_{r(3)})-
    \Phi(\mathscr{C}_{2}-C_{r(3)})
    \right)
    \\
    &+&\left. \left(1 - 
    \Phi(\mathscr{C}_{3}-C_{r(4)})
    \right)\right],
\end{eqnarray}
where $\Phi$ is the cumulative distribution function of a Gaussian with zero mean and standard deviation $\sigma_C$:
\begin{equation}
\Phi(C) = \frac{1}{\sqrt{2 \pi} \sigma_C} \int_{-\infty}^{C} dC' \exp\left(-\frac{C'^2}{2\sigma_C^2}\right).
\end{equation}
The fidelity is bounded by $0.25\leq\mathcal F\leq1$.

The assignment fidelity as the function of detuning is plotted in Fig.~\ref{fig:setup}(d), as the blue curve, for our example parameter set (see caption), and for amplifier noise strength $\sigma_C = 1\, \mathrm{fF}$.
For these parameters, the fidelity approaches unity for certain detuning values.
Furthermore, as expected, the fidelity correlates with the capacitance contrast (gray curve in Fig.~\ref{fig:setup}(d)).

\begin{figure}[t]
    \centering
    \includegraphics[width=1\linewidth]{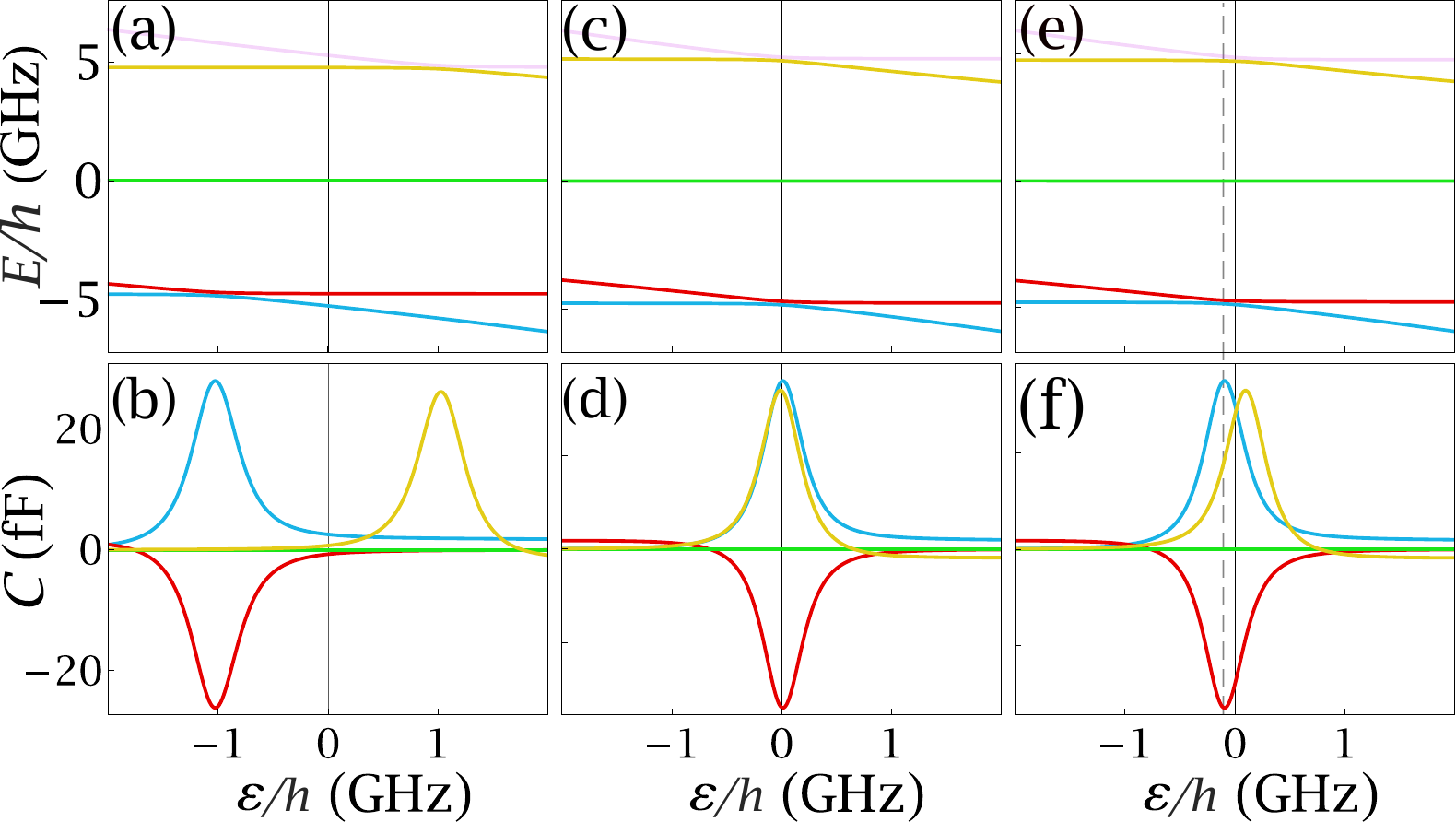}
    \caption{\textbf{Maximizing the capacitance contrast for a fixed magnetic-field configuration.} Energy levels (a,c,e) and quantum capacitances (b,d,f) corresponding to the four lowest-energy states, as functions of detuning $\varepsilon$, for different values of tunneling energy $t_0$. 
    Magnetic-field parameters: $(B_{a\parallel},B_{a\perp
    },B_s) = (-2.5, 5, 171.8)$ mT. 
    (a,b) Misaligned lower and upper singlet-triplet anticrossings, at $t_0/h = 3.72$ GHz. 
    (c,d) Aligned anticrossings, at $t_0/h = t_\mathrm{align}/h = 3.37$ GHz. 
    (e,f) Slightly misaligned anticrossings, at $t_0/h = 3.41$ GHz, which maximizes the capacitance contrast at the detuning value marked by the dashed vertical line.}
    \label{fig:spec-cq-6panel}
\end{figure}

\subsection{Maximizing the capacitance contrast}
\label{sec:perturbation}

In our model, the capacitance contrast $\Delta C$ depends on the gate-tunable parameters $\varepsilon$ and  $t_0$, as well as the magnetic-field parameters $B_s$, $B_{a\parallel}$, and $B_{a\perp}$.
In this subsection, we provide a procedure to maximize the capacitance contrast.
To illustrate this procedure, we start from the limiting case where the local magnetic fields in the two dots are similar; that is, $B_{a\parallel},B_{a\perp} \ll B_s$.

We have seen already in Fig.~\ref{fig:setup}(b) that the misalignment of the local magnetic fields in the two dots, which imply a nonzero value of the parameter $B_{a\perp}$, opens up two singlet-triplet anticrossings in the detuning-dependent energy spectrum. 
We refer to those as the lower (between $E_1$ and $E_2$) and upper (between $E_4$ and $E_5$) anticrossings.
We argue that the capacitance contrast is maximized when the lower and upper singlet-triplet anticrossings are positioned at slightly different detuning values.
To see this, consider the three energy diagrams and three quantum capacitance diagrams shown in Fig.~\ref{fig:spec-cq-6panel}.

Fig.~\ref{fig:spec-cq-6panel}(a) shows a generic situation, when the singlet-triplet anticrossings are  misaligned, that is, their detuning positions are far from each other.
In this case, the capacitance contrast is poor for any value of $\varepsilon$: if $\varepsilon$ is tuned to, say, the upper (lower) singlet-triplet anticrossing, then $C_2\approx C_3\approx0$ ($C_3\approx C_4\approx0$) and hence $\Delta C \approx 0$.

It is possible to fine-tune the tunneling energy $t_0$ such that the lower and upper singlet-triplet anticrossings are aligned, that is, are positioned at the same detuning value $\varepsilon = 0$.
In the limit of negligibly small magnetic-field difference $B_a \ll B_s$, this is reached if
\begin{equation}
\label{eq:alignment_condition}
t_0 = t_\mathrm{align} \equiv \frac{1}{\sqrt{2}} g \muB B_s.
\end{equation}
With that choice, shown in Fig.~\ref{fig:spec-cq-6panel}(c,d), the two positive-valued capacitance peaks (blue $C_1$ and yellow $C_4$) corresponding to the two singlet-triplet bonding states, overlap, yielding, again, a weak capacitance contrast $\Delta C \approx 0$.

However, a maximum of capacitance contrast can be found starting from the latter situation, by a small change $\delta t_\mathrm{opt}$ of the tunneling energy $t_0$; that is, setting $t_0 = t_\mathrm{align} + \delta t_\mathrm{opt}$.
E.g., a slight increase of $t_0$, as shown in Fig.~\ref{fig:spec-cq-6panel}(e,f), displaces the blue (yellow) capacitance peak of $C_1$ ($C_4$) to the left (right).
Then, for certain values of the detuning $\varepsilon$, all four capacitances become distinguishable; this happens, for example, at the small negative $\varepsilon$ value indicated by the dashed vertical line in Fig.~\ref{fig:spec-cq-6panel}(e,f).

In what follows, we provide an analytical, perturbative formula for the tunneling energy change $\delta t_\mathrm{opt}$ to be applied with respect to Eq.~\eqref{eq:alignment_condition}, in order to maximize the capacitance contrast. 
Here, we consider the case $0 < \delta t_\mathrm{opt}$.
We claim that the capacitance contrast is maximized for tunneling energy $t_0 = t_\mathrm{align} + \delta t_\mathrm{opt}$ and detuning $\varepsilon = \varepsilon_\mathrm{opt}$, with 
\begin{equation}
\label{eq:optimaldeltat}
\delta t_\mathrm{opt} \approx \frac{\sqrt{2^{2/3}-1}}{2 \sqrt{2}} \, 
g \muB B_{a\perp} \approx
0.27 \, g \muB B_{a\perp},
\end{equation}
and
\begin{equation}
\label{eq:optimaldeltaepsilon}
\varepsilon_\mathrm{opt} \approx - 2 \sqrt{2} \, \delta t_\mathrm{opt}
\approx - 0.77\, g \muB B_{a\perp}.
\end{equation}

To provide a perturbative derivation of Eqs.~\eqref{eq:optimaldeltat} and \eqref{eq:optimaldeltaepsilon}, we split the Hamiltonian $H$ of Eq.~\eqref{eq:H5} to an unperturbed part and a perturbation, the latter being defined as the antisymmetric Zeeman term:
\begin{subequations}
    \begin{align}
        H &= H_0 + H_1, \\
        H_0 &= \mqty(-\varepsilon & \sqrt 2\, t_0 & 0 & 0 & 0\\
                           \sqrt2\, t_0 & 0 & 0 & 0 & 0\\
                           0 & 0 & 0 & 0 & 0\\
                           0 & 0 & 0 & g \muB B_s & 0\\
                           0 & 0 & 0 & 0 & -g \muB B_s),\label{eq:hunperturbed}\\
        H_1 &= g \muB \mqty(0 & 0 & 0 & 0 & 0\\
                           0 & 0 &  B_{a\parallel} & -\frac{B_{a\perp}}{\sqrt 2} & \frac{B_{a\perp}}{\sqrt 2}\\
                           0 & B_{a\parallel} & 0 & 0 & 0\\
                           0 & -\frac{B_{a\perp}}{\sqrt 2} & 0 & 0 & 0\\
                           0 & \frac{B_{a\perp}}{\sqrt 2} & 0 & 0 & 0).\label{eq:hperturbation}
    \end{align}
\end{subequations}

From $H_0$, we can determine the detuning value corresponding to the lower singlet-triplet anticrossing. 
We denote the bonding (antibonding) singlet energy eigenstate as $\ket{\mathcal{S}_-(\varepsilon)}$ ($\ket{\mathcal{S}_+(\varepsilon)}$) of $H_0$.
The corresponding energy eigenvalues of $H_0$ are 
\begin{equation}
\label{eq:singletenergies}
E_{\mathcal{S}_{\pm}}(\varepsilon) = 
- \frac \varepsilon 2 \pm
\frac 1 2
\sqrt{\varepsilon^2+8 t^2_0}
\approx
- \frac \varepsilon 2 \pm \sqrt{2}t_0.
\end{equation}
Here, the approximation is valid if $\varepsilon \ll t_0$.
Furthermore, the polarized triplets have energy eigenvalues
\begin{equation}
\label{eq:tripletenergies}
E_{\mathcal{T}_\pm} = \pm g \muB B_s.    
\end{equation}
Eqs.~\eqref{eq:singletenergies} and \eqref{eq:tripletenergies} imply that the detuning values corresponding to the lower and upper singlet-triplet anticrossings, determined by the equations $E_{\mathcal{S}_-}(\varepsilon) = E_{\mathcal{T}_-}$ and
$E_{\mathcal{S}_+}(\varepsilon) = E_{\mathcal{T}_+}$, respectively, coincide at $\varepsilon = 0$ if Eq.~\eqref{eq:alignment_condition} holds.

The detuning-dependent capacitances $C_1(\varepsilon)$, $C_2(\varepsilon)$ and $C_4(\varepsilon)$ can be expressed analytically from the corresponding $2\times 2$ effective Hamiltonians.
We do this at $t_0 = t_\mathrm{align}$ and $\varepsilon \approx 0$.
Close to the lower singlet-triplet anticrossing, the effective Hamiltonian is obtained by projecting $H$ onto the subspace of $\ket{\mathcal{S}_-(\varepsilon = 0)}$ and $\ket{\mathcal{T}_-}$, yielding:
\begin{equation}
\label{eq:hefflower}
    H_\mathrm{lower} = \left(\begin{array}{cc}
    - \varepsilon/2 & g\muB B_{a\perp}/2 \\
     g\muB B_{a\perp}/2 & 0
    \end{array}
    \right),
\end{equation}
where we used the approximation of Eq.~\eqref{eq:singletenergies}, and omitted an irrelevant term proportional to the unit matrix.

We express a simple approximative formula for the quantum capacitances $C_1$ and $C_2$ from Eq.~\eqref{eq:hefflower} as follows. 
We express the bonding and antibonding eigenvectors 
$\chi_b = (\chi_{b,\mathcal{S}},\chi_{b,\mathcal{T}})^\mathrm{T}$ 
and
$\chi_a = (\chi_{a,\mathcal{S}},\chi_{a,\mathcal{T}})^\mathrm{T}$ of Eq.~\eqref{eq:hefflower}.
Then, we approximate the energy eigenstate $\ket{\psi_1}$ as 
\begin{eqnarray}
\ket{\psi_1(\varepsilon)} \approx
\chi_{b,\mathcal{S}}(\varepsilon) \ket{\mathcal{S}_-(\varepsilon=0)}
+
\chi_{b,\mathcal{T}}(\varepsilon)
\ket{\mathcal{T}_-},
\end{eqnarray}
and $\ket{\psi_2}$ in a similar fashion.
Finally, we evaluate the quantum capacitance using Eq.~\eqref{eq:quantumcapacitance}, which yields:
\begin{equation}
\label{eq:c1epsilon}
    C_1(\varepsilon) = 
    \frac{e^2}{8}
    \frac{g^2\muB^2 B^2_{a\perp}}{\left(
    \varepsilon^2/4 + 
    g^2\muB^2 B^2_{a\perp}
    \right)^{3/2}}.
\end{equation}
Furthermore, it holds that  $C_2(\varepsilon) = - C_1(\varepsilon)$
and 
$C_4(\varepsilon) = C_1(\varepsilon)$.
We will denote the peak value of the capacitance curve $C_1(\varepsilon)$ as $C_\mathrm{peak} = e^2/(8 g \muB B_{a\perp})$.

With a small change $\delta t$ of the tunneling energy, i.e., setting $t_0 = t_\mathrm{align}+\delta t$, the capacitance peaks of $C_1$ and $C_4$ are displaced into opposite directions along the detuning axis:
\begin{equation}\label{eq:C1_smallBperp}
C_1(\varepsilon;\delta t) = 
    \frac{e^2}{8}
    \frac{g^2\muB^2 B^2_{a\perp}}{\left(
    \left(\frac{\varepsilon}{2} + \sqrt{2} \delta t\right)^2 + 
    g^2\muB^2 B^2_{a\perp}
    \right)^{3/2}},
\end{equation}
and
\begin{equation}\label{eq:C4_smallBperp}
C_4(\varepsilon;\delta t) = 
    \frac{e^2}{8}
    \frac{g^2\muB^2 B^2_{a\perp}}{\left(
    \left(\frac{\varepsilon}{2} - \sqrt{2} \delta t\right)^2 + 
    g^2\muB^2 B^2_{a\perp}
    \right)^{3/2}}.
\end{equation}
To maximize the capacitance contrast, i.e., to reach $\Delta C = C_\mathrm{peak}/2$, parameters $\varepsilon$ and $\delta t$ should be fine-tuned such that $C_1(\varepsilon,\delta t) = C_\mathrm{peak}$ and $C_4(\varepsilon,\delta t) = C_\mathrm{peak}/2$. 
These two relations imply the results in Eqs.~\eqref{eq:optimaldeltat} and \eqref{eq:optimaldeltaepsilon}.

\begin{figure}[t]
\centering
\includegraphics[width=1\linewidth]{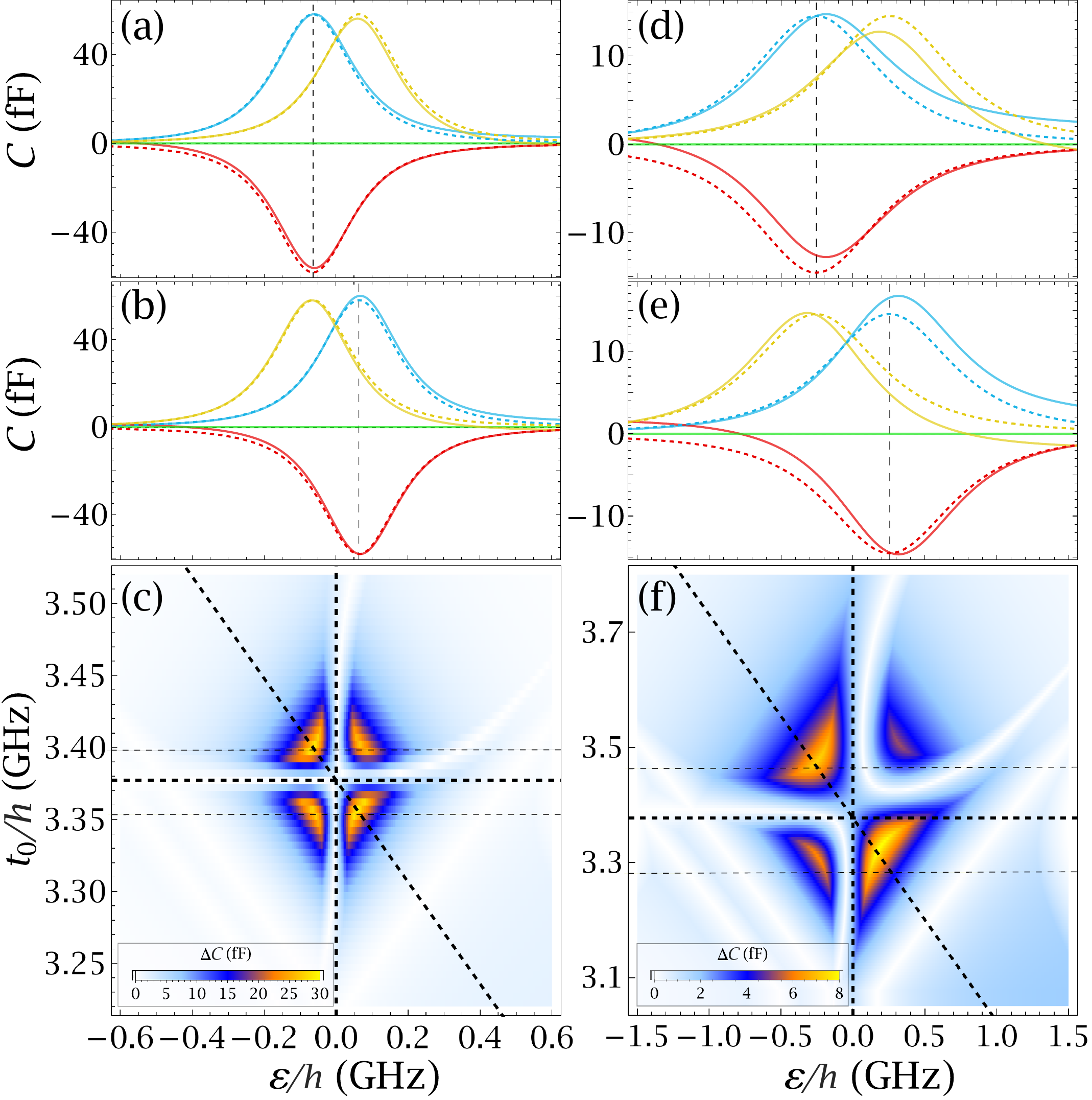}
\caption{\textbf{Maximizing the capacitance contrast in the detuning-tunneling parameter plane.}
(a,b) Detuning dependence of the quantum capacitances of the four states to be discriminated, for magnetic-field parameters
$(B_{a\parallel},B_{a\perp}, B_s) = (-2.5, 3.0, 171.8)$ mT.
Tunneling energy:
(a) $t_0/h = 3.399$ GHz and
(b) $t_0/h =3.354$ GHz,
corresponding to $t_0 = t_\mathrm{align} \pm \delta t$, respectively. 
Solid: exact numerical result.
Dashed: perturbative result (e.g., Eq.~\eqref{eq:C1_smallBperp}).
(c) Detuning and tunneling dependence of the quantum capacitance contrast $\Delta C$, for the same magnetic-field parameters.
Exact, numerical results are shown.
Panel (a) [(b)] corresponds to the upper [lower] horizontal cut of panel (c) denoted by the thin dashed line.
Horizontal thick dashed line indicates $t_\mathrm{align}/h = 3.377$ GHz.
Diagonal thick dashed line is the line containing 
$(\varepsilon,t_0) = (0,t_\mathrm{align})$
and the perturbatively determined location of optimal capacitance contrast
$(\varepsilon,t_0) = (\varepsilon_\mathrm{opt},t_\mathrm{align}+\delta t_\mathrm{opt})$ (see Eqs.~\eqref{eq:optimaldeltat},\eqref{eq:optimaldeltaepsilon}).
(d,e,f) Analogous to (a,b,c), with fourfold increased $B_{a\perp}$, i.e., 
$(B_{a\parallel},B_{a\perp}, B_s) = (-2.5, 12.0, 171.8)$ mT.
Tunneling energy: (d) $t_0/h = 3.47$ GHz, (e) $t_0/h = 3.29$ GHz.
Increasing $B_{a\perp}$ from (d) to (f) reduces the number of optimal readout positions (capacitance contrast maxima) from 4 to 2.
}
\label{fig:cq-perturbation-density}
\end{figure}
As shown in Fig.~\ref{fig:cq-perturbation-density}(a,b), the above perturbative results (dashed lines) are in good agreement with numerically exact results (solid lines) in the limit of small $B_a$ (see caption for parameters).
In Fig.~\ref{fig:cq-perturbation-density}(d,e), where the perturbation $B_{a\perp}$ is increased by a factor of four, the exact results (data points) start to deviate from the perturbative results (solid lines), illustrating the limitations of the above perturbative description.

The capacitance contrast $\Delta C(\varepsilon,t_0)$ as the function of detuning and tunneling energy shows a characteristic quatrefoil pattern in the perturbative regime, as shown in Fig.~\ref{fig:cq-perturbation-density}(c).
This density plot shows exact numerical results.
The quatrefoil pattern signals that in the vicinity of the anticrossing alignment point $(\varepsilon,t_0) = (0,t_\mathrm{align})$, there are four equivalent optimal readout positions maximizing the capacitance contrast.

Moving from Fig.~\ref{fig:cq-perturbation-density}(a,b,c) to Fig.~\ref{fig:cq-perturbation-density}(d,e,f), we increase $B_{a\perp}$ from 3 mT to 12 mT. 
In Fig.~\ref{fig:cq-perturbation-density}(f), we observe that the fourfold symmetry of the quatrefoil pattern breaks down, leaving only two optimal readout positions instead of four.
This characteristic change is in accord with the fact that the peak height of $C_1(\varepsilon)$ is greater than the peak height of $C_4(\varepsilon)$ in the numerical data in Fig.~\ref{fig:cq-perturbation-density}(d,e).
In turn, this is due to the fact that the quantum capacitance $C_1$ ($C_4$) has a contribution from the quantum capacitance of the bonding (antibonding) singlet state $\ket{\mathcal{S}_-}$
($\ket{\mathcal{S}_+}$), which is positive (negative), and hence increases (decreases) the dominant positive contribution induced by the singlet-triplet anticrossing matrix element.

To summarize, in this subsection we described a procedure to optimize the gate-tunable detuning and tunneling energy parameters to maximize the capacitance contrast. 
We found that for weak $B_{a\perp}$, there are four equivalent, locally optimal readout points in the $(\varepsilon,t_0)$ plane, reducing to two equivalent optima for stronger $B_{a\perp}$.
We also provided a  perturbative procedure to derive analytical formulas for those optimal readout points.

\section{Optimizing four-state discrimination in the presence of relaxation}
\label{sec:relaxation}

\begin{figure}[t]
\centering
\includegraphics[width=1\linewidth]{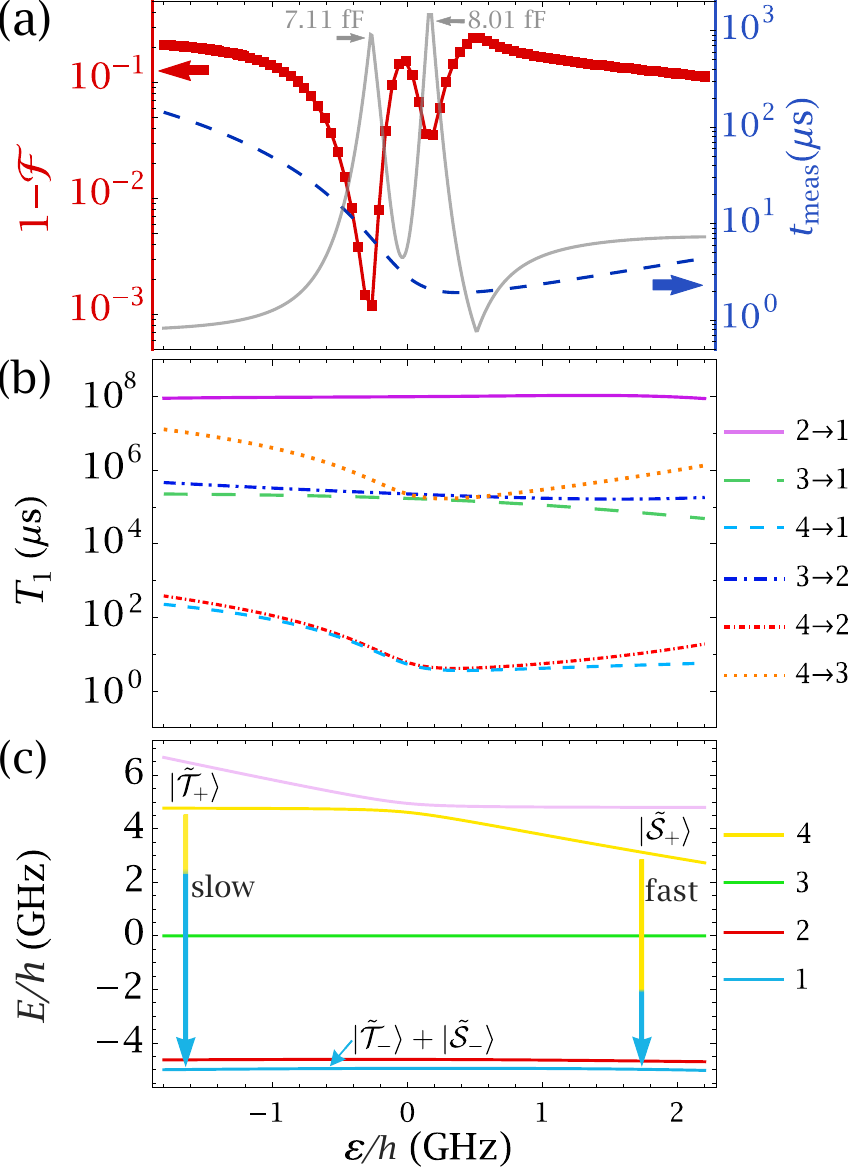}
\caption{\textbf{Assignment infidelity in the presence of phonon-mediated relaxation and amplifier noise.} 
(a) Red (left y axis): Average assignment infidelity $1-\mathcal{F}$ as function of detuning $\varepsilon$, 
along the diagonal dashed line of Fig.~\ref{fig:cq-perturbation-density}(f),
at an amplifier noise level $c = 3.136\, \mathrm{fF} \sqrt{\mu \mathrm{s}}$ (see Eq.~\eqref{eq:sigmactmeas}).
Blue dashed (right y axis): measurement time $t_\textrm{meas}$, set as the minimum decay time  at 50 mK via Eq.~\eqref{eq:tmeas}. 
Gray solid line (y axis not shown): capacitance contrast $\Delta C$.
Two infidelity dips appear, correlated with the two capacitance-contrast peaks.
Difference of the depths of the two infidelity dips is explained by the difference of the relaxation rates. 
(b) Phonon-mediated downhill relaxation times  among the four computational basis states at zero temperature. 
The $4\to 1$ relaxation paths are denoted by arrows; the slow (fast) path requires spin flip (no spin flip).
(c) Two-electron DQD energy spectrum.
The common x axis of the three panels corresponds to the detuning $\varepsilon$ along the diagonal dashed line of Fig.~\ref{fig:cq-perturbation-density}(f).
}
\label{fig:amplifier_noise}
\end{figure}

For state-of-the-art readout of solid-state qubits, data acquisition time for a single shot ranges from a few tens of nanoseconds to a few milliseconds. 
Within that measurement duration, environment-induced relaxation processes have a strong influence on the quality of the readout \cite{Barthel2009SingleShot,DanjouPRBDispersiveReadout,Vigneau2023}.

In fact, the relaxation time $T_1$ is often regarded as the maximal reasonable data acquisition time for single-shot readout, since beyond $T_1$, the information in the pre-measurement qubit state is mostly erased. 
Therefore, we compute relaxation times between the relevant states for 4SD, and quantify how these relaxation processes influence the assignment fidelity of 4SD.

In Sec.~\ref{sec:amplifier-noise}, amplifier noise was taken into account as a random component of the measured quantum capacitance value.
Here, we also take into account how the standard deviation $\sigma_C$ of that random component depends on the measurement duration. We will use the following simple formula to describe this dependence: 
\begin{equation}
\label{eq:sigmactmeas}
\sigma_C = \frac{\sqrt{k_\mathrm B T_n R}}{\kappa V_0\sqrt{t_\textrm{meas}}} = \frac{c}{\sqrt{t_\textrm{meas}}}.
\end{equation}
where the second equality defines the constant $c$. 
Parameters in Eq.~\eqref{eq:sigmactmeas}, and their values to be used in the following, are as follows (see Ref.~\cite{Vahid2020} and Eqs.~(2) and (5) therein for more details): $\kappa=0.6 \, \mathrm{fF}^{-1}$ is the coefficient converting quantum capacitance to reflectance, $V_0 = 3 \, \mu \mathrm{V}$ is the voltage amplitude of the probe signal on gate $R$, $T_n = 4 \, \mathrm{K}$ is the effective noise temperature of the amplifier, and $R = 577 \, \textrm{k}\Omega$ is the the resistance of the tank circuit.
Together, these yield $c=3.136$ fF$/\sqrt{\mathrm{\mu s}}$.

Here, we estimate the assignment infidelity for a given set of model parameters. 
Our procedure is as follows. 
First, recall that the parameters in our DQD model are $\varepsilon$, $t_0$, $B_s$, $B_{a\parallel}$, $B_\perp$. 
Additional parameters influencing the relaxation process are the noise strength constant $c$ of Eq.~\eqref{eq:sigmactmeas}, the interdot distance $d$, and the temperature $T$.
We compute the zero-temperature and finite-temperature phonon-induced relaxation times $T_{1,j\to j'}$ among the 4 lowest-energy eigenstates $j,j' \in \{1,2,3,4\}$ of a silicon DQD with interdot distance $d = 100$ nm, see details in App.~\ref{appendix:phonons}. 
We also compute the total decay time of each state via
\begin{equation}
\label{eq:decaytimes}
   T^{-1}_j = \sum_{j' \neq j} T^{-1}_{1, j\to j'}.
\end{equation}
Then we set the measurement time 
equal to the shortest of the decay times:
\begin{equation}
\label{eq:tmeas}
t_\textrm{meas}
=
\min_{j \in \{1,2,3,4\}} T_{j}.
\end{equation}
Finally, we evaluate the assignment infidelity $1 - \mathcal{F}$ from Eq.~\eqref{eq:average_assignment_fidelity}, such that $\sigma_C$ is evaluated as $\sigma_C(t_\textrm{meas})$ using Eq.~\eqref{eq:sigmactmeas}, to obtain the assignment fidelity.

Results for the assignment infidelity are shown as the red data points in Fig.~\ref{fig:amplifier_noise}(a), as the function of detuning $\varepsilon$ along the diagonal dashed line of Fig.~\ref{fig:cq-perturbation-density}(f).
Whereas in Fig.~\ref{fig:cq-perturbation-density}(f), the capacitance contrast peaks at positive and negative detuning values are approximately symmetric, in Fig.~\ref{fig:amplifier_noise}(c), the two infidelity dips are highly asymmetric, with the negative-detuning dip showing much better infidelity.

We explain this asymmetry of the infidelity dips in Fig.~\ref{fig:amplifier_noise}(a) by the detuning asymmetry of the zero-temperature $4 \to 1$ relaxation time. 
The zero-temperature downhill relaxation times calculated from Eq.~\eqref{eq:spin-qubit-relaxation-rate} are shown in Fig.~\ref{fig:amplifier_noise}(b). 
As shown in Fig.~\ref{fig:amplifier_noise}(b), $T_{1,4\to1}$ is the shortest of the six relaxation times, for all detuning values considered in the figure; hence, at low temperature, this is the relaxation time that controls the measurement time $t_\mathrm{meas}$ of Eq.~\eqref{eq:tmeas}.
As the measurement time is longer for the negative-detuning readout position than for the positive-detuning readout position, the negative-detuning position yields lower assignment infidelity and hence favored.

Finally, we explain the dominance of the relaxation time $T_{1,4\to 1}$ and its detuning asymmetry.
To facilitate this explanation, we plot the detuning-dependent energy spectrum of the five energy eigenstates in Fig.~\ref{fig:amplifier_noise}(c).
The relaxation processes $4 \to 1$ and $4\to 2$ are at least three orders of magnitude faster than the other processes. 
These short times scales are explained by the large energy gaps $\sim 2 \sqrt{2} t_0$ between the initial and final states, and the  large available phonon density of states at the corresponding phonon frequencies.
Furthermore, even though in Fig.~\ref{fig:amplifier_noise}(c) the energy gap $E_4-E_1$ is slightly asymmetric about $\varepsilon=0$, this has little effect on the asymmetry of the relaxation times in Fig.~\ref{fig:amplifier_noise}(b). 
Instead, that relaxation-time asymmetry is due to spin selection rules (which are quantified by the matrix elements in Fermi's Golden Rule in \eqref{eq:fermi_golden_rule}). 
On the one hand, for negative detuning, the initial state $4$ is mostly $\ket{\mathcal{T}_+}$, whereas the final state $1$ is a balanced hybrid of $\ket{\mathcal{T}_-}$ and the bonding singlet $\ket{\mathcal{S}_-}$; hence relaxation requires spin flip and hence it is slow. 
On the other hand, for positive detuning, the initial state $4$ is mostly the antibonding singlet $\ket{\mathcal{S}_+}$, whereas the final state is the same as above, hence relaxation does not require spin flip, and hence is fast.

In this section, we focused on the effect of phonon-mediated relaxation processes on the assignment infidelity of 4SD. Relaxation provides a natural upper bound to the measurement time, and hence a lower bound on the assignment infidelity.
We have quantified these bounds focusing on a silicon-based double quantum dot with a micromagnet, and identified a single optimal readout position where the spectral and spin structure of the energy eigenstates ensure long relaxation times and hence low assignment infidelity.

\section{Discussion}
\label{sec:discussion}

Our results highlight an opportunity to perform 4SD of two-electron states in a DQD with gate reflectometry. 
In this section, we outline potential ways to refine our analysis further, enabling more precise modeling and more accurate 4SD. 

For this work, we assumed that the reflectance measurement in the gate-reflectometry setup including the DQD reveals the quantum capacitances of the spin states.
In reality, the relation between the reflectance and the quantum capacitance may have nontrivial ingredients, and modeling those might be important in certain settings. 

For example, the reflectance can be influenced by coherent or dissipative dynamical effects in the DQD, such as overdrive effects  in terms of Landau-Zener transitions \cite{Vahid2020}, or tunneling capacitance, Sisyphus resistance \cite{Mizuta2017,Esterli2019}, Hermes admittance \cite{Peri2024Unified}, etc, which go beyond the adiabatic dynamics that underlies the concept of quantum capacitance. 
These effects can be made small in the setup we considered, if the resonator's time period is chosen well below the above-microsecond relaxation times (see Fig.~\ref{fig:amplifier_noise}(b)), and the probe signal is weak enough.

Interestingly, these dynamical effects not only lead to readout error, but also impose measurement back-action that reduces the quantum non-demolition (QND) character of the readout process. 
In practical terms \cite{Yoneda2020QNDSi}, a measurement is described as QND, if, after measuring a certain outcome, an immediate subsequent measurement yields the same outcome with unit probability.
The dynamical effects described above reduce the QND character of the readout, since they induce dynamics that redistributes the population among the energy eigenstates during the course of the measurement, hence lead to a finite probability that two subsequent measurements yield different outcomes.

We also note that in this work, for simplicity, we assumed a trivial lever arm structure for the DQD, and hence our estimates for the quantum capacitances are upper bounds, which should be weighted by the lever arm squared \cite{Mizuta2017,Vigneau2023}, to obtain realistic numbers.

A further feature of real devices is charge noise \cite{Vahid2020}, whose main effect can be incorporated into our model as detuning jitter.
When too strong, charge noise can lead to an increased readout error via broadening and suppressing the detuning dependence of the capacitance contrast peak (shown as gray in Fig.~\ref{fig:setup}(d)).

Although we postpone the quantification of the above-described mechanisms for future work, we nevertheless provide a practical numerical example on how the quantum capacitance is revealed in a reflectometry experiment. 
For this, we take the classical tank circuit setup and its parameters from Ref.~\cite{AhmedRFSensing}.
Furthermore, we use the quantum capacitance values 
$C_i \in \{C_1,C_2,C_3,C_4\}$
from Fig.~\ref{fig:cq-perturbation-density}(d) at the dashed-line cut, and scale them with a realistic gate lever arm \cite{Vigneau2023} value $\alpha = 0.2$, yielding rescaled quantum capacitance values 
$C'_i= \alpha^2 C_i = (0.58,-0.5,0,0.29) \, \mathrm{fF}$. 
With the circuit parameters of Ref.~\cite{AhmedRFSensing}, using a probe frequency that is resonant with the circuit at vanishing quantum capacitance, we obtain reflectance phase shift values of 
$\varphi_i=(-90^{\circ},82^{\circ},0^{\circ},-62^{\circ})$, respectively.

The principle of the 4SD procedure we proposed here is material agnostic, as long as either spin-orbit interaction or an inhomogeneous magnetic field provides the spectral anticrossings that are exploited by the method.
For n-type silicon DQDs, the case we focused on, a low valley splitting might be a hindrance in general for qubit functionality \cite{Zwanenburg2013SiQuantElec}.
For 4SD however, the readout point is in the vicinity of the (1,1)-(0,2) charge transition, hence the valley excited states are expected to play a much less significant adversarial role than in charge sensing, the latter being performed deep in the (0,2) configuration. 

We also anticipate that careful design and characterization of the classical tank circuit of the reflectometry setup is an important requirement to discriminate the four quantum capacitance values. 
For example, if the tank circuit resonance of the reflectance $\Gamma(C)$ (see, e.g., Fig.~(7) of \cite{Vahid2020}) is narrower than the capacitance contrast $\Delta C$, then discriminating all four states becomes difficult.
The recently developed co-simulation tool for hybrid quantum-classical circuits \cite{Peri2025MicrowaveSim} might be especially useful for optimizing the tank circuit design for 4SD.

Another interesting extension of our analysis is to consider advanced inference methods for 4SD. 
In our work, we assumed that the quantum capacitance is inferred from the time-averaged reflectance signal. 
Such an approach disregards soft information present in the time trace of the reflectance, and hence can be improved upon by time-trace analysis, for the price of a computational overhead in data processing.
One approach is based on hidden Markov models, which has already been studied \cite{Spethmann} and implemented \cite{Laine2025dispersiveHMM} in the context of spin qubit readout.

We also briefly discuss single-shot and multi-shot applications of the 4SD we proposed. 
Throughout this paper, we used the single-shot assignment fidelity to characterize 4SD, since single-shot readout is required for general quantum algorithms.
However, there are special quantum information protocols, ranging from simple quantum state tomography \cite{RyanTomography,FilippPRL2009} to complex digital quantum simulation algorithms \cite{KimUtility}, where the goal is to infer expectation values of observables for the final state of a quantum circuit.
For that purpose, multi-shot readout can be utilized, and provides superior performance if the single-shot readout fidelity is too low \cite{RyanTomography,DAnjouSoft}. 
In the context of 4SD via gate reflectometry, multi-shot readout means that the same circuit is performed multiple times, the reflectance time traces for those shots are averaged, and the populations of the computational basis states are estimated from the shot-averaged time traces.
This multi-shot version of 4SD is expected to be a powerful gadget for quantum information experiments with semiconductor spin qubits.
We also note similarly to the 4SD procedure we describe, simultaneous multi-shot readout of two qubits have been demonstrated with superconducting qubits \cite{FilippPRL2009,DiCarloNature2009,ChowPRA2010} and nitrogen-vacancy centers in diamond \cite{JingyanHePRApp2024}.

In which cases do we expect that 4SD will replace usual ancilla-based readout of semiconductor spin qubits?
Of course, it will do so in any setup where 4SD is superior to ancilla-based readout in all figures of merits (e.g., readout error, QND-ness, back-action, etc).
Consider hence a setup where 4SD is inferior in some, or all, of those figures of merits. 
Even then, 4SD is superior in at least one figure of merit: the number of available data qubits. 
For example, having a 100-qubit array, the experimentalists can decide if they operate it the conventional way, as a 50-qubit quantum processing unit -- i.e., they use 50 qubits as data qubits and 50 qubits as ancillas for readout --, or they operate it as a 100-qubit quantum processing unit, making use of 4SD.
It is a plausible scenario that the experimentalists want to demonstrate digital quantum simulation \cite{KimUtility} for a problem (e.g., dynamics of an interacting quantum spin system) that can be classically simulated with 50 spins, but cannot be classically simulated with 100 spins.
In such cases, the experimentalist will likely use 4SD as a readout method, even if this bring some overhead in terms of characterizing and mitigating readout error.

\section{Conclusions}
\label{sec:conclusions}

We proposed, modeled, and analyzed an experimental procedure, which enables the discrimination of the four spin states of two electrons in a double quantum dot in a single measurement.
The procedure discriminates the quantum capacitances of the four states using gate reflectometry.
We described a recipe to maximize the contrast between the quantum capacitances of the four basis states by an appropriate choice of detuning and interdot tunneling parameters. 
Focusing on silicon double quantum dots equipped with a micromagnet, we quantified how amplifier noise and phonon-mediated relaxation limits the readout infidelity.
Our results highlight a realistic opportunity to mitigate the overhead of readout ancilla qubits in a spin-based quantum computer.

\begin{acknowledgments}
We thank B. Kolok, M. Rimbach-Russ and M.~Urdampilleta for constructive feedback on the project, and 
B.~Undseth for useful input on the micromagnetic field configurations.
This research was supported by the Ministry of Culture and Innovation (KIM) and the National Research, Development and Innovation Office (NKFIH) within the Quantum Information National Laboratory of Hungary (Grant No.~2022-2.1.1-NL-2022-00004), by the HUN-REN Hungarian Research Network through the HUN-REN-BME-BCE Quantum Technology Research Group, and by the European Union within the Horizon Europe research and innovation programme via the IGNITE, ONCHIPS, and QLSI2 projects.
\end{acknowledgments}

\section*{Code Availability}

The code used to obtain the results and generate the figures is available at~\cite{zenodoSen}.

\appendix

\section{Computing quantum capacitance of numerically obtained energy eigenstates without numerical differentiation}
\label{appendix:capacitance-derivation}

In the main text, we expressed the quantum capacitances of the DQD energy eigenstates in Eq.~\eqref{eq:quantumcapacitance}.
The formula suggests that if the energy eigenstates are computed numerically, then numerical differentiation is required to evaluate the quantum capacitance. 
This is not the case.
In this appendix, we provide a formula to calculate quantum capacitance from first-order time independent perturbation theory, i.e., in a way that does not require the finite-difference numerical evaluation of any derivative. 
In all of our figures, we used this formula to compute the numerically exact values of quantum capacitances.

We claim that the quantum capacitance formula of Eq.~\eqref{eq:quantumcapacitance} is equivalent to the following formula: 
\begin{equation}\label{eq:cq_d2E_equation}
    C_j = -2 e^2 \sum_{k\neq j}
    \frac{|\braket{\psi_k | n_\mathrm R | \psi_j}|^2}{E_j - E_k},
\end{equation}
or, if the effective lever arm $\alpha$ of the right dot \cite{Mizuta2017}
is not 1, then
\begin{equation}\label{eq:cq_d2E_equation_withleverarm}
    C_j = -2 \alpha^2 e^2 \sum_{k\neq j}
    \frac{|\braket{\psi_k | n_\mathrm R | \psi_j}|^2}{E_j - E_k}.
\end{equation}
In our figures, we have used Eq.~\eqref{eq:cq_d2E_equation} to evaluate the quantum capacitance of the energy eigenstates.

One way to derive Eq.~\eqref{eq:cq_d2E_equation} is as follows.
Consider the perturbative problem of $\tilde{H} = H+H'$, where $H' = \nu n_R$ is the contribution of a small change of the right plunger gate voltage to the Hamiltonian.
Then, Eq.~\eqref{eq:quantumcapacitance} can be expressed as
\begin{equation}
C_j = e^2 \lim_{\nu \to 0}\frac{
\braket{\tilde{\psi}_j | n_R | \tilde{\psi}_{j}} - 
\braket{\psi_j | n_R | \psi_{j}}
}{\nu},
\end{equation}
where the tilde marks the perturbed wave functions.
Expressing the perturbed wave functions using first-order perturbation theory, followed by straightforward simplifications, yields Eq.~\eqref{eq:cq_d2E_equation}.

\begin{figure}[t]
    \centering
    \includegraphics[width=0.9\linewidth]{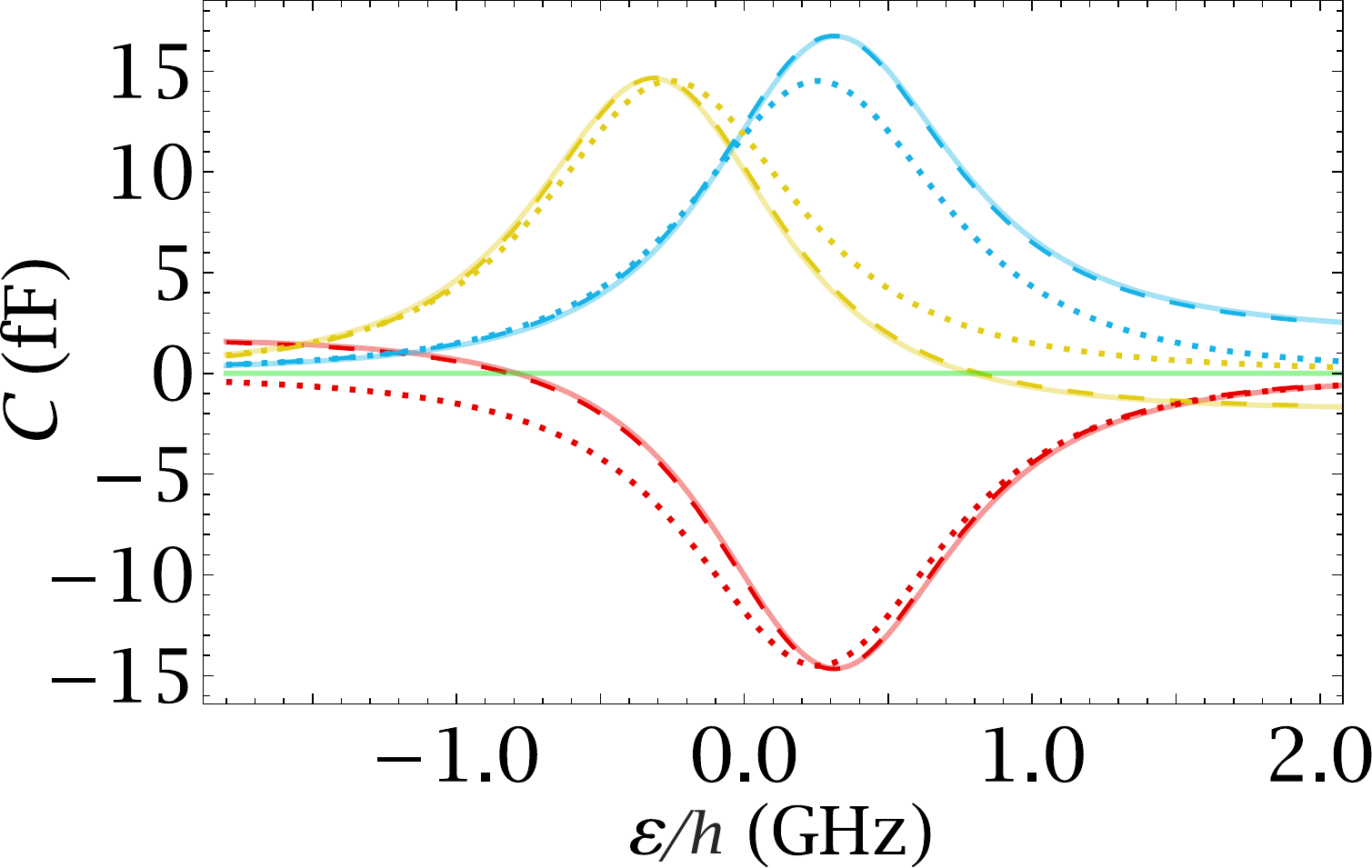}
    \caption{\textbf{Comparison of numerically exact and perturbatively calculated quantum capacitances.}
    Dotted: quantum capacitances derived from the simplified perturbative approach of the main text (Eqs.~\eqref{eq:C1_smallBperp},
    \eqref{eq:C4_smallBperp}).
    Dashed: result of the more accurate perturbative method described in App.~\ref{app:refined}    (Eqs.~\eqref{eq:psi_tilde_1},\eqref{eq:psi_tilde_2}).
    Solid: numerically exact results (Eq.~\eqref{eq:cq_d2E_equation}). 
    Dashed and solid lines overlap, illustrating that the refined perturbative approach provides an accurate description of the numerically exact results.
    Parameters are the same as in Fig.~\ref{fig:cq-perturbation-density}(e).}
    \label{figapp:exact_perturbation_capacitances}
\end{figure}

\section{A more accurate perturbative description of the quantum capacitances at the singlet-triplet anticrossings}
\label{app:refined}

In Sec.~\ref{sec:perturbation}, we used a perturbative description of the quantum capacitances of the four relevant energy eigenstates in the vicinities of the lower and upper singlet-triplet anticrossings. 
Here, we argue that this perturbative description can be refined and made more accurate,  for the price of slightly more complicated analytical formulas. 
Here, we describe this refined procedure, and although we omit those more complicated formulas, we do compare the results with the simpler perturbative results and the exact numerical results.
In short, this refined method takes into account the quantum capacitances of the bonding and antibonding singlet states, which are neglected by the simplified description of Sec.~\ref{sec:perturbation}.

We illustrate the procedure for the lower singlet-triplet anticrossing. 
The derivation refines that in Sec.~\ref{sec:perturbation} in the step preceding Eq.~\eqref{eq:hefflower}; now we project onto the subspace of $\ket{\mathcal{S}_-(\varepsilon)}$ and $\ket{\mathcal{T}_-}$, i.e,. we account for the detuning dependence of the bonding singlet state. 
\begin{equation}
\label{eq:hefflower_exact}
   H_\mathrm{lower} = \mqty(E_{\mathcal{S}_-}(\varepsilon) && 
   \frac{g\mu_B B_{a\perp}}{\sqrt 2}\sin(\theta/2)
   \\ 
   \frac{g\mu_B B_{a\perp}}{\sqrt 2}\sin (\theta/2) && 
   -g\muB B_s), 
\end{equation}
where $\tan(\theta)=2\sqrt 2 t_0/\varepsilon$. 
We denote the bonding and antibonding eigenvectors of this $H_\mathrm{lower}$ as
$\chi_b(\varepsilon) = (\chi_{b,\mathcal{S}}(\varepsilon),\chi_{b,\mathcal{T}}(\varepsilon))^\mathrm{T}$ 
and
$\chi_a = (\chi_{a,\mathcal{S}}(\varepsilon),\chi_{a,\mathcal{T}}(\varepsilon))^\mathrm{T}$ of Eq.~\eqref{eq:hefflower_exact}.
Then, the energy eigenstates are expressed as
\begin{subequations}\label{eq:quantum_capacitance_varied_basis}
\begin{align}
\ket{\psi_1(\varepsilon)} &=
\chi_{b,\mathcal{S}}(\varepsilon) \ket{\mathcal{S}_-(\varepsilon)}
+
\chi_{b,\mathcal{T}}(\varepsilon)
\ket{\mathcal{T}_-}\label{eq:psi_tilde_1},\\
\ket{\psi_2(\varepsilon)} &=
\chi_{a,\mathcal{S}}(\varepsilon) \ket{\mathcal{S}_-(\varepsilon)}
+
\chi_{a,\mathcal{T}}(\varepsilon)
\ket{\mathcal{T}_-}\label{eq:psi_tilde_2}.
\end{align}
\end{subequations}
From these wave function, one can derive analytical results for the quantum capacitances $C_1$ and $C_2$ via Eq.~\eqref{eq:quantumcapacitance}.
Furthermore, $C_4$ can be described in an analogous way, focusing on the upper singlet-triplet anticrossing.

In Fig.~\ref{figapp:exact_perturbation_capacitances}, we compare the quantum capacitances derived from the simplified perturbative approach of the main text 
(Eqs.~\eqref{eq:C1_smallBperp},\eqref{eq:C4_smallBperp}, dotted lines) with the more accurate perturbative method described in this section 
(Eqs.~\eqref{eq:psi_tilde_1},\eqref{eq:psi_tilde_2}, dashed lines) and from the numerically exact method 
(Eq.~\eqref{eq:cq_d2E_equation}, solid lines). 
The dashed and solid lines overlap, illustrating that the refined perturbative approach introduced in this section provides an accurate description of the numerically exact results.

\section{Phonon-induced relaxation rates in a silicon double quantum dot}
\label{appendix:phonons}

As discussed in Sec.~\ref{sec:relaxation} of the main text, phonon-mediated relaxation is an important mechanism limiting the achievable readout fidelity. 
Here, following earlier studies \cite{TaylorPRB2007,DanonPRB2013,Boross2016Nanotech,Kornich2018RelaxationSiGe}, we describe the details of computing the relaxation times.
First, we focus on the case of zero temperature, and later we generalize to finite temperatures.
In our model, we consider the effect of bulk phonons in silicon.
The results obtained here are applied in Sec.~\ref{sec:relaxation}.

In the parameter range we consider, the energy splittings between the electronic states correspond to the frequency range of low-energy acoustic phonons, hence we focus on those.
There are three polarization modes, $\lambda \in \{\mathrm{L}, \mathrm{T1}, \mathrm{T2}\}$, where the polarization indices L and T refer to longitudinal and transverse, respectively.
The dispersion relations of these low-frequency phonons are linear, $\omega = v_\lambda q$, characterized by their sound velocities $v_\lambda$.

Focusing on n-type quantum dots in silicon, we use the Herring-Vogt electron-phonon Hamiltonian ~\cite{Herring1957,Tahan2014Relaxation,Boross2016ValleySi,Boross2016Nanotech,Kornich2018RelaxationSiGe}, describing the deformation-potential mechanism:
\begin{equation}\label{eq:hamiltonian_e_ph}
    H_\mathrm{eph} = \Xi_d \Tr(\epsilon) + \Xi_u \epsilon_{zz}.
\end{equation}
Here, $\Xi_d$ is the dilational deformation potential, whereas $\Xi_u$ is the uniaxial deformation potential,.
Note that for quantum dots, defined in a layer perpendicular to the 001 direction, the electronic wave function is localized in the $z$ and $\bar z$ valleys of the silicon band structure. 
The Hamiltonian of Eq.~\eqref{eq:hamiltonian_e_ph} contains only the diagonal part of the strain tensor $\epsilon$, which is expressed as:
\begin{equation}\label{eq:straintensor}
    \epsilon_{jj} = \mi \sqrt{\frac{\hbar}{2\rho\, V}}\sum_{\vec q, \lambda}\frac{\left(\boldsymbol{e}_{\vec q, \lambda}\right)_j q_j}{\sqrt{v_\lambda\, q}}e^{i\vec q\cdot\vec r}
    \left(a_{\vec q, \lambda} + a^\dagger_{-\vec q, \lambda}
    \right)
\end{equation}
where $j\in(x,y,z)$, and $e_{\vec q,\lambda}$ is the polarization vector of the polarization mode $\lambda$. 

To calculate the zero-temperature relaxation rate between two electronic energy levels, one at higher energy ($\ket{e}$) and the other at lower energy ($\ket{g}$), we use Fermi's Golden Rule:
\begin{equation}\label{eq:fermi_golden_rule}
    \Gamma_0 = \frac{2\pi}{\hbar}\sum_{\vec q, \lambda}\big|\bra{g,\vec q\lambda}H_\text{eph}\ket{e,0}\big|^2\delta(\Delta E-\hbar  v_\lambda q),
\end{equation}
where $\Delta E$ is the energy difference between the two levels.

\subsection{Relaxation of a charge qubit}

First, we consider phonon-assisted relaxation of a charge qubit defined in a double quantum dot.
We assume that the relaxation is caused by the interplay of (i) the interdot motion of the electron caused by the phonon-induced on-site energy difference between the two dots (ii) and the different magnetic fields in the two dots.

This mechanism is captured by assuming strongly localized, Dirac-delta-type local basis states on each dot \cite{Boross2016Nanotech}.
We take a reference frame such that the center of dot L is at the origin, whereas the center of dot R is at $\vec r_d = (d,0,0)$.
Then, the localized basis functions are approximated as
$\braket{{\vec{r}} | L} = \sqrt{\delta(\vec r)}$
and
$\braket{\vec r | R} = \sqrt{\delta(\vec r-\vec r_d)}$.
The charge-qubit Hamiltonian in the left-right basis defined above reads:
\begin{equation}
\label{eq:hchargequbit}
    H_\mathrm{c} = \frac{\varepsilon}{2} \sigma_z
    + \frac{\Delta}{2} \sigma_x, 
\end{equation}
where $\varepsilon$ is the detuning between the on-site energies, $\Delta$ is the interdot tunneling energy, and $\sigma_x$ and $\sigma_z$ are Pauli matrices acting on the left-right basis defined above.

We restrict the electronic degrees of freedom of the electron-phonon interaction Hamiltonian $H_\mathrm{eph}$ to the two above-defined, localized charge-qubit basis states. 
That is, we define the projection 
$P = \dyad{L}+\dyad{R}$, 
and use the projected Hamiltonian
\begin{equation}
H_\mathrm{eph,c}=PH_\mathrm{eph}P.
\end{equation}
This restriction implies that $H_\mathrm{eph,c}$ has the same form as the right hand side of Eq.~\eqref{eq:hamiltonian_e_ph}, but in the expression \eqref{eq:straintensor} for the strain tensor, we replace 
\begin{equation}\label{eq:ExpOfStrain-down-projection}
e^{i \vec q \cdot \vec r} \mapsto \ket{L}\bra{L} + e^{i q_x d} \ket{R}\bra{R}.
\end{equation}
Correspondingly, in Fermi's Golden Rule \eqref{eq:fermi_golden_rule}, we replace $H_\mathrm{eph} \mapsto H_\mathrm{eph,c}$, and reintepret the electronic states $\ket{e}$ and $\ket{g}$ as the two-component eigenvectors of $H_\mathrm{c}$ in Eq.~\eqref{eq:hchargequbit}.

The right hand side of Fermi's Golden Rule \eqref{eq:fermi_golden_rule} is a sum of 3 contributions from the polarization modes.
Without the loss of generality, the polarization vectors $\vec e_{\vec q,\lambda}$ for the three polarizations can be chosen as follows:
\begin{subequations}\label{eq:polarization-vectors}
\begin{align}
    \vec e_{\vec q,\mathrm L} &= \frac{\vec q}{q} = \mqty(\sin \theta\cos \phi\\ \sin\theta\sin\phi\\ \cos\theta),\label{eq:polarization-vectors-L}\\
    \vec e_{\vec q,\mathrm T1} &= \mqty(\cos \theta\cos \phi\\ \cos\theta\sin\phi\\-\sin\theta),\label{eq:polarization-vectors-T1}\\
    \vec e_{\vec q,\mathrm T2} &= \mqty(-\sin \phi\\ \cos\phi\\0)\label{eq:polarization-vectors-T2}
\end{align}
\end{subequations}
Here, the momentum vector has been parametrized with spherical coordinates $\vec q = q (\sin \theta \cos \phi, \sin\theta \sin \phi, \cos \theta)$.

The transverse modes do not contribute to dilational term in Eq.~\eqref{eq:hamiltonian_e_ph}, as the polarization vector $\vec e_{\vec q,\mathrm{T}}\perp\vec q$, such that $\text{Tr}(\epsilon)=0$. The polarization vector in Eq.~\eqref{eq:polarization-vectors-T2} does not contribute to the second term in Eq.~\eqref{eq:hamiltonian_e_ph} either. 

First, wecalculate the contributions from the longitudinal phonons.
The diagonal elements of strain tensor, expressed in the orbital basis restricted to $\ket{L}$ and $\ket{R}$, read as follows:
\begin{equation}\label{eq:longitudinal-strain-tensor}
\epsilon^\mathrm{L}_{\mathrm{c},jj} = \mi\,
\sqrt{\frac{\hbar}{2\rho V v_\mathrm{L}}}
\sum_{\boldsymbol{q}}
\frac{q_j^2}{q^{3/2}}
\left(1-e^{iq_x d}\right)
\frac{\sigma_z}{2}
    \left(a_{\vec q, \mathrm{L}} + a^\dagger_{-\vec q, \mathrm{L}}
    \right),
\end{equation}
where we used Eqs.~\eqref{eq:straintensor}, \eqref{eq:ExpOfStrain-down-projection}, and \eqref{eq:polarization-vectors-L}. Here, we omitted a term proportional to $\sigma_0$, describing an equal on-site energy shift on the two dots, which does not contribute to relaxation. 

The contribution in Fermi's Golden Rule from the longitudinal phonon due to the dilational deformation potential contains the following matrix element
\begin{eqnarray}
M^{\mathrm{L}}_{\vec q, d} 
&=& \mel{g,\vec q L}{H_\mathrm{eph,c}}{e,0}
\nonumber
\\
&=&
\mi\, \Xi_d\sqrt{\frac{\hbar}{8\rho V v_\mathrm{L}}}\sum_{q'}\sqrt{q'}\big(1-e^{\mi q'_x d}\big)
\nonumber
\\
&\times&
\bra{g,\vec q L}\sigma_z\left(a_{\vec q', \mathrm{L}} + a^\dagger_{-\vec q', \mathrm{L}}\right)\ket{e,0}.
\end{eqnarray}    
Evaluating the matrix element yields:    
\begin{equation}
M^{\mathrm{L}}_{\vec q, d} 
=
\mi\,\Xi_d\sqrt{\frac{\hbar}{8\rho V v_\mathrm{L}}}\mel{g}{\sigma_z}{e}\sum_{\vec q'}\delta_{\vec{q,-q'}}\sqrt{q'}\big(1-e^{\mi q'_x d}\big).
\end{equation}
Finally, evaluating the sum yields:
\begin{equation}
\label{eq:longitudinal-dilational-matrix-element}
M^{\mathrm{L}}_{\vec q, d} 
=
\mi\,\Xi_d\sqrt{\frac{\hbar}{8\rho V v_\mathrm{L}}}\mel{g}{\sigma_z}{e}\sqrt{q}\big(1-e^{-\mi q_x d}\big)
\end{equation}
where $\ket g,\ket e$ are two-component complex vectors as introduced below Eq.~\eqref{eq:ExpOfStrain-down-projection}.

The uniaxial deformation potential term has only the z-valley contribution from the strain tensor,
\begin{equation}
    \epsilon^{\mathrm L}_{\mathrm{c},zz} = \mi\, \sqrt{\frac{\hbar}{8\rho V v_\mathrm{L}}}
    \sum_{\vec q}
    \frac{q^2_z}
    {q^{3/2}}\big(1-e^{\mi q_x d}\big)\sigma_z\left(a_{\vec q, \mathrm{L}} + a^\dagger_{-\vec q, \mathrm{L}}\right)
\end{equation}
and the corresponding matrix element in Fermi's Golden Rule reads:
\begin{equation}\label{eq:longitudinal-uniaxial-matrix-element}
   M^{\mathrm{L}}_{\vec q,u} =\mi\,\Xi_u\sqrt{\frac{\hbar}{8\rho V v_\mathrm{L}}}\mel{g}{\sigma_z}{e}\sqrt{q}\cos^2\theta\big(1-e^{-\mi q_x d}\big).
\end{equation}

As discussed above, only the transverse phonons with polarization vector $e_{\vec q,\mathrm T1}$ contribute to relaxation, and they do so only through the uniaxial deformation potential. 
The strain tensor element for the transverse phonons reads:
\begin{equation}\label{eq:transverse-strain-tensor}
    \epsilon^{\mathrm T}_{c,zz}=\mi\,\sqrt{\frac{\hbar}{8\rho V v_{\mathrm T}}}\frac{q_z\,\vec e_{z,\mathrm T1}}{\sqrt q}\left(1-e^{iq_x d}\right)\sigma_z\left(a_{\vec q, \mathrm{T}} + a^\dagger_{-\vec q, \mathrm{T}}\right),
\end{equation}
and the corresponding matrix element appearing in Fermi's Golden Rule reads:
\begin{equation}\label{eq:transverse-matrix-element}
    M^{\mathrm T}_{\vec q,u} =-\mi\, \Xi_u\sqrt{\frac{\hbar}{8\rho V v_\mathrm{T}}}\mel{g}{\sigma_z}{e}\sqrt{q}\sin\theta\cos\theta\big(1-e^{-\mi q_x d}\big).
\end{equation}
The zero-temperature relaxation rate calculated using Fermi's Golden Rule \eqref{eq:fermi_golden_rule}, applying Eqs.~\eqref{eq:longitudinal-dilational-matrix-element}, ~\eqref{eq:longitudinal-uniaxial-matrix-element} for the the longitudinal phonons,
and 
Eq. ~\eqref{eq:transverse-matrix-element}
for the transverse phonons.
We convert the sum over the phonon wavenumber into an integral in spherical coordinates. 
The rates associated to the emission of longitudinal and transversal phonons, respectively, are evaluated as follows:
\begin{widetext}
\begin{subequations}\label{eq:charge-relaxation-rate}
    \begin{align}
        \Gamma_{0,\mathrm L} &= \frac{2\pi}{\hbar}\sum_{\vec q}| M^\mathrm{L}_{\vec q}|^2\delta(\Delta E - \hbar v_{\mathrm L}q)=
        \frac{2\pi}{\hbar}\sum_{\vec q}
        |
        M^\mathrm{L}_{\vec q,d} + 
        M^\mathrm{L}_{\vec q,u}
        |^2
        \delta(\Delta E - \hbar v_{\mathrm L}q)\nonumber\allowdisplaybreaks\\
        &= \frac{2\pi}{\hbar}\frac{\hbar}{8\rho V v_\mathrm{L}}|\mel{g}{\sigma_z}{e}|^2\sum_{\vec q}|(1 - e^{-\mi q_x d})|^2 q(\Xi^2_d + 2\Xi_u\Xi_d \cos^2\theta + \Xi^2_u \cos^4\theta)\delta(\Delta E - \hbar v_{\mathrm L} q)\nonumber\allowdisplaybreaks\\
        &=  \frac{\pi}{4\rho V v_\mathrm{L}}|\mel{g}{\sigma_z}{e}|^2\frac{V}{(2\pi)^3}\int^\infty_0 dq q^3 \int^{2\pi}_0d\phi\int^\pi_0 \sin \theta d\theta\, 2(1-\cos (q_x d))(\Xi^2_d + 2\Xi_u\Xi_d \cos^2\theta + \Xi^2_u \cos^4\theta)\delta(\Delta E - \hbar v_{\mathrm L} q)\nonumber\allowdisplaybreaks\\
        &= \frac{1}{16\pi^2\rho v_\mathrm{L}}|\mel{g}{\sigma_z}{e}|^2\frac{(\Delta E)^3}{\hbar^4 v^4_\mathrm{L}}\int^{2\pi}_0d\phi\int^\pi_0 d\theta\,\sin \theta \Bigg(1-\cos (\frac{\Delta E}{\hbar v_\mathrm L}\, d\, \sin \theta \cos \phi)\Bigg)(\Xi^2_d + 2\Xi_u\Xi_d \cos^2\theta + \Xi^2_u \cos^4\theta)\nonumber\allowdisplaybreaks\\
        &=\frac{(\Delta E)^3}{16\pi^2\hbar^4 v^5_\mathrm{L}\rho}|\mel{g}{\sigma_z}{e}|^2\big(\Xi^2_d I_0 + 2\Xi_d\Xi_u I_2 + \Xi^2_u I_4\big),\label{eq:charge-relaxation-rate-longitudinal}\\
        &\vspace{15pt}\nonumber\allowdisplaybreaks\\
        \Gamma_{0,\mathrm T} &= \frac{2\pi}{\hbar}\sum_{\vec q}| M^{\mathrm{T}}_{\vec q}|^2\delta(\Delta E - \hbar v_{\mathrm T}q)\nonumber\allowdisplaybreaks\\ 
        &=\frac{2\pi}{\hbar}\frac{\hbar\, \Xi^2_u}{8\rho V v_\mathrm T}|\mel{g}{\sigma_z}{e}|^2\frac{V}{(2\pi)^3}\int^\infty_0 dq\, q^3 \int^{2\pi}_0 d\phi \int^\pi_0 d\theta \sin^3\theta \cos^2\theta (1 - \cos(q\,d \sin\theta\cos\phi))\delta(\Delta E - \hbar v_{\mathrm T} q)\nonumber\allowdisplaybreaks\\
        &=\frac{\Xi^2_u}{16\, \pi^2\rho v_\mathrm T}|\mel{g}{\sigma_z}{e}|^2\frac{(\Delta E)^3}{\hbar^4 v^4_\mathrm T} \int^{2\pi}_0 d\phi \int^\pi_0 d\theta \sin^3\theta \cos^2\theta \Bigg(1 - \cos(\frac{\Delta E}{\hbar v_\mathrm T}\,d \sin\theta\cos\phi)\Bigg)\nonumber\allowdisplaybreaks\\
        &=\frac{\Xi^2_u\,(\Delta E)^3}{16\, \pi^2\hbar^4 v^5_\mathrm T\rho }|\mel{g}{\sigma_z}{e}|^2 J\label{eq:charge-relaxation-rate-transverse},\\
        &\vspace{15pt}\nonumber\allowdisplaybreaks\\
        \Gamma_0 &= \Gamma_{0,\mathrm L} + \Gamma_{0, \mathrm T}.\label{eq:charge-relaxation-rate-total}
\end{align}
\end{subequations}
Here, we have introduced the following integrals:
\begin{equation}\label{eq:spherical-integration_I_n}
\begin{aligned}
    I_n &= \int^{2\pi}_0d\phi\int^{\pi}_0d\theta \sin \theta (\cos \theta)^n\times\Bigg(1-\cos (\frac{\Delta E}{\hbar v_\mathrm L}\, d\, \sin \theta \cos \phi)\Bigg),\\
    J &=\int^{2\pi}_0 d\phi \int^\pi_0 d\theta \sin^3\theta \cos^2\theta\times \Bigg(1 - \cos(\frac{\Delta E}{\hbar v_\mathrm T}\,d \sin\theta\cos\phi)\Bigg).
\end{aligned}
\end{equation}
\end{widetext}

\subsection{Relaxation between two-electron energy eigenstates}

To describe the two-electron relaxation times shown in Fig.~\ref{fig:amplifier_noise}, we perform calculations analogous to those presented in the previous subsection. 
We denote the zero temperature downhill relaxation rate as $\Gamma^{j,k}_0 = T^{-1}_{1,k \to j}$ for each pair $j,k$ of energy eigenstates with $E_j < E_k$. 
We project our electron-phonon Hamiltonian to the five basis states, whose spatial structure we assume to be Dirac-delta-like. 
This implies that we apply the following replacement to the plane wave factor in the strain tensor of Eq.~\eqref{eq:straintensor}:
\begin{equation}
    e^{\mi \vec q\cdot\vec r} \mapsto 
    \left(e^{i q_x d} 
    - 1\right)
    \dyad{\mathcal S},
\end{equation}
where we omitted an irrelevant term proportional to the unit matrix. 

With this replacement, we can recycle our results of Eq.~\eqref{eq:charge-relaxation-rate} obtained for the charge qubit, for our current goal of describing two-electron relaxation.
In fact, the matrix element in Eq.~\eqref{eq:charge-relaxation-rate} should be replaced as 
\begin{equation}
\braket{g | \sigma_z |e}
\mapsto 
2 \braket{g | \mathcal S} 
\braket{\mathcal S | e}
\equiv 
2 \braket{\psi_j | \mathcal S} 
\braket{\mathcal S | \psi_k}
=: 2 A_{j,k}.
\end{equation}
Therefore, the zero-temperature relaxation rate is given by
\begin{equation}\label{eq:spin-qubit-relaxation-rate}
\begin{aligned}
    \Gamma^{j,k}_0 &= \frac{(\Delta E_{j,k})^3|A_{j,k}|^2}{4\pi^2\hbar^4\rho}\\
    &\hspace{20pt}\times\Big(\frac{\Xi^2_d I^{j,k}_0 + 2\Xi_d\Xi_u I^{j,k}_2 + \Xi^2_u I^{j,k}_4}{v^5_\mathrm L} + \frac{\Xi^2_u\, J^{j,k}}{v^5_\mathrm T}\Big),
\end{aligned}
\end{equation}
where the $j,k$ upper index in the spherical integrals specifies the energy gap $\Delta E_{j,k} = E_k - E_j$ between the  two states. 

We used the following parameter values \cite{Boross2016ValleySi} to calculate the relaxation rates: $d=100$ nm, $\Xi_u=8.77$ eV, $\Xi_d = 5$ eV, $v_\mathrm L = 9330$ m/s, $v_\mathrm T = 5420$ m/s and $\rho = 2330$ kg/m$^3$. The matrix elements $A_{j,k}$ and the energy gaps $\Delta E_{j,k}$ are calculated numerically for Fig.~\ref{fig:amplifier_noise}.
To obtain finite-temperature downhill and uphill relaxation times, required to evaluate the decay times and the measurement time according to Eqs.~\eqref{eq:decaytimes} and \eqref{eq:tmeas}, we used the Bose-Einstein factors
\begin{equation}
n^{j,k}_p = (e^{\Delta E_{j,k}/k_\mathrm B T} - 1)^{-1}
\end{equation}
to obtain 
\begin{eqnarray}
T^{-1}_{1,k\to j} = 
\Gamma^{j,k}_0(1 + n^{j,k}_p)
\end{eqnarray}
for the downhill case, and
\begin{eqnarray}
T^{-1}_{1,j\to k} = 
\Gamma^{j,k}_0 n^{j,k}_p
\end{eqnarray}
for the uphill case.

\bibliography{references}

@misc{zenodoSen,
  author       = {Sen, Aritra},
  title        = {{Codes to support the findings in "Four-state
                   discrimination for a pair of spin qubits via gate
                   reflectometry"
                  }},
  month        = jun,
  year         = 2026,
  publisher    = {Zenodo},
  doi          = {10.5281/zenodo.20513171},
  url          = {https://doi.org/10.5281/zenodo.20513171},
}

@article{DAnjouSoft,
  title = {Soft Decoding of a Qubit Readout Apparatus},
  author = {D'Anjou, B. and Coish, W. A.},
  journal = {Phys. Rev. Lett.},
  volume = {113},
  issue = {23},
  pages = {230402},
  numpages = {5},
  year = {2014},
  month = {Dec},
  publisher = {American Physical Society},
  doi = {10.1103/PhysRevLett.113.230402},
  url = {https://link.aps.org/doi/10.1103/PhysRevLett.113.230402}
}

@article{KimUtility,
	author = {Kim, Youngseok and Eddins, Andrew and Anand, Sajant and Wei, Ken Xuan and van den Berg, Ewout and Rosenblatt, Sami and Nayfeh, Hasan and Wu, Yantao and Zaletel, Michael and Temme, Kristan and Kandala, Abhinav},
	journal = {Nature},
	number = {7965},
	pages = {500--505},
	title = {Evidence for the utility of quantum computing before fault tolerance},
	volume = {618},
	year = {2023}}

@article{RyanTomography,
  title = {Tomography via correlation of noisy measurement records},
  author = {Ryan, Colm A. and Johnson, Blake R. and Gambetta, Jay M. and Chow, Jerry M. and da Silva, Marcus P. and Dial, Oliver E. and Ohki, Thomas A.},
  journal = {Phys. Rev. A},
  volume = {91},
  issue = {2},
  pages = {022118},
  numpages = {7},
  year = {2015},
  month = {Feb},
  publisher = {American Physical Society},
  doi = {10.1103/PhysRevA.91.022118},
  url = {https://link.aps.org/doi/10.1103/PhysRevA.91.022118}
}

@article{JingyanHePRApp2024,
  title = {Direct readout of a nitrogen-vacancy hybrid-spin quantum register in diamond by analysis of photon arrival time},
  author = {He, Jingyan and Tian, Yu and Hu, Zhiyi and Ye, Runchuan and Wang, Xiangyu and Lu, Dawei and Xu, Nanyang},
  journal = {Phys. Rev. Appl.},
  volume = {21},
  issue = {5},
  pages = {054041},
  numpages = {12},
  year = {2024},
  month = {May},
  publisher = {American Physical Society},
  doi = {10.1103/PhysRevApplied.21.054041},
  url = {https://link.aps.org/doi/10.1103/PhysRevApplied.21.054041}
}

@article{DiCarloNature2009,
   author = {L. DiCarlo and J. M. Chow and J. M. Gambetta and Lev S. Bishop and B. R. Johnson and D. I. Schuster and J. Majer and A. Blais and L. Frunzio and S. M. Girvin and R. J. Schoelkopf},
   doi = {10.1038/nature08121},
   issn = {0028-0836},
   issue = {7252},
   journal = {Nature},
   month = {7},
   pages = {240-244},
   title = {Demonstration of two-qubit algorithms with a superconducting quantum processor},
   volume = {460},
   year = {2009}
}

@article{ChowPRA2010,
  title = {Detecting highly entangled states with a joint qubit readout},
  author = {Chow, J. M. and DiCarlo, L. and Gambetta, J. M. and Nunnenkamp, A. and Bishop, Lev S. and Frunzio, L. and Devoret, M. H. and Girvin, S. M. and Schoelkopf, R. J.},
  journal = {Phys. Rev. A},
  volume = {81},
  issue = {6},
  pages = {062325},
  numpages = {8},
  year = {2010},
  month = {Jun},
  publisher = {American Physical Society},
  doi = {10.1103/PhysRevA.81.062325},
  url = {https://link.aps.org/doi/10.1103/PhysRevA.81.062325}
}

@article{FilippPRL2009,
  title = {Two-Qubit State Tomography Using a Joint Dispersive Readout},
  author = {Filipp, S. and Maurer, P. and Leek, P. J. and Baur, M. and Bianchetti, R. and Fink, J. M. and G\"oppl, M. and Steffen, L. and Gambetta, J. M. and Blais, A. and Wallraff, A.},
  journal = {Phys. Rev. Lett.},
  volume = {102},
  issue = {20},
  pages = {200402},
  numpages = {4},
  year = {2009},
  month = {May},
  publisher = {American Physical Society},
  doi = {10.1103/PhysRevLett.102.200402},
  url = {https://link.aps.org/doi/10.1103/PhysRevLett.102.200402}
}

@article{TaylorPRB2007,
  title = {Relaxation, dephasing, and quantum control of electron spins in double quantum dots},
  author = {Taylor, J. M. and Petta, J. R. and Johnson, A. C. and Yacoby, A. and Marcus, C. M. and Lukin, M. D.},
  journal = {Phys. Rev. B},
  volume = {76},
  issue = {3},
  pages = {035315},
  numpages = {17},
  year = {2007},
  month = {Jul},
  publisher = {American Physical Society},
  doi = {10.1103/PhysRevB.76.035315},
  url = {https://link.aps.org/doi/10.1103/PhysRevB.76.035315}
}

@article{DanonPRB2013,
  title = {Spin-flip phonon-mediated charge relaxation in double quantum dots},
  author = {Danon, J.},
  journal = {Phys. Rev. B},
  volume = {88},
  issue = {7},
  pages = {075306},
  numpages = {8},
  year = {2013},
  month = {Aug},
  publisher = {American Physical Society},
  doi = {10.1103/PhysRevB.88.075306},
  url = {https://link.aps.org/doi/10.1103/PhysRevB.88.075306}
}

@article{HsiaoExciton,
  title = {Exciton Transport in a Germanium Quantum Dot Ladder},
  author = {Hsiao, T.-K. and Cova Fari\~na, P. and Oosterhout, S. D. and Jirovec, D. and Zhang, X. and van Diepen, C. J. and Lawrie, W. I. L. and Wang, C.-A. and Sammak, A. and Scappucci, G. and Veldhorst, M. and Demler, E. and Vandersypen, L. M. K.},
  journal = {Phys. Rev. X},
  volume = {14},
  issue = {1},
  pages = {011048},
  numpages = {17},
  year = {2024},
  month = {Mar},
  publisher = {American Physical Society},
  doi = {10.1103/PhysRevX.14.011048},
  url = {https://link.aps.org/doi/10.1103/PhysRevX.14.011048}
}

@article{vanDiepenSimulation,
  title = {Quantum Simulation of Antiferromagnetic Heisenberg Chain with Gate-Defined Quantum Dots},
  author = {van Diepen, C. J. and Hsiao, T.-K. and Mukhopadhyay, U. and Reichl, C. and Wegscheider, W. and Vandersypen, L. M. K.},
  journal = {Phys. Rev. X},
  volume = {11},
  issue = {4},
  pages = {041025},
  numpages = {15},
  year = {2021},
  month = {Nov},
  publisher = {American Physical Society},
  doi = {10.1103/PhysRevX.11.041025},
  url = {https://link.aps.org/doi/10.1103/PhysRevX.11.041025}
}

@article{DehollainNagaoka,
	author = {Dehollain, J. P. and Mukhopadhyay, U. and Michal, V. P. and Wang, Y. and Wunsch, B. and Reichl, C. and Wegscheider, W. and Rudner, M. S. and Demler, E. and Vandersypen, L. M. K.},
	date = {2020/03/01},
	doi = {10.1038/s41586-020-2051-0},
	id = {Dehollain2020},
	isbn = {1476-4687},
	journal = {Nature},
	number = {7800},
	pages = {528--533},
	title = {Nagaoka ferromagnetism observed in a quantum dot plaquette},
	url = {https://doi.org/10.1038/s41586-020-2051-0},
	volume = {579},
	year = {2020}}

@article{ScherublCommsPhys2019,
   author = {Zoltán Scherübl and András Pályi and György Frank and István Endre Lukács and Gergő Fülöp and Bálint Fülöp and Jesper Nygård and Kenji Watanabe and Takashi Taniguchi and Gergely Zaránd and Szabolcs Csonka},
   doi = {10.1038/s42005-019-0200-2},
   issn = {2399-3650},
   issue = {1},
   journal = {Communications Physics},
   month = {9},
   pages = {108},
   title = {Observation of spin–orbit coupling induced {W}eyl points in a two-electron double quantum dot},
   volume = {2},
   url = {https://www.nature.com/articles/s42005-019-0200-2},
   year = {2019}
}

@article{PfundPRL2007,
  title = {Suppression of Spin Relaxation in an {InAs} Nanowire Double Quantum Dot},
  author = {Pfund, A. and Shorubalko, I. and Ensslin, K. and Leturcq, R.},
  journal = {Phys. Rev. Lett.},
  volume = {99},
  issue = {3},
  pages = {036801},
  numpages = {4},
  year = {2007},
  month = {Jul},
  publisher = {American Physical Society},
  doi = {10.1103/PhysRevLett.99.036801},
  url = {https://link.aps.org/doi/10.1103/PhysRevLett.99.036801}
}

@article{PioroLadriereNatPhys2008,
	author = {Pioro-Ladri{\`e}re, M. and Obata, T. and Tokura, Y. and Shin, Y. -S. and Kubo, T. and Yoshida, K. and Taniyama, T. and Tarucha, S.},
	journal = {Nature Physics},
	number = {10},
	pages = {776--779},
	title = {Electrically driven single-electron spin resonance in a slanting Zeeman field},
    doi = {10.1038/nphys1053},
    url = {https://doi.org/10.1038/nphys1053},
	volume = {4},
	year = {2008}}

@article{BuehlerAPL2005,
    author = {Buehler, T. M. and Reilly, D. J. and Starrett, R. P. and Greentree, Andrew D. and Hamilton, A. R. and Dzurak, A. S. and Clark, R. G.},
    title = {Single-shot readout with the radio-frequency single-electron transistor in the presence of charge noise},
    journal = {Applied Physics Letters},
    volume = {86},
    number = {14},
    pages = {143117},
    year = {2005},
    month = {04},
    issn = {0003-6951},
    doi = {10.1063/1.1897423},
    url = {https://doi.org/10.1063/1.1897423}
}

@article{SchleserAPL2004,
    author = {Schleser, R. and Ruh, E. and Ihn, T. and Ensslin, K. and Driscoll, D. C. and Gossard, A. C.},
    title = {Time-resolved detection of individual electrons in a quantum dot},
    journal = {Applied Physics Letters},
    volume = {85},
    number = {11},
    pages = {2005-2007},
    year = {2004},
    month = {09},
    issn = {0003-6951},
    doi = {10.1063/1.1784875},
    url = {https://doi.org/10.1063/1.1784875}
}

@article{CollessPRL2013,
  title = {Dispersive Readout of a Few-Electron Double Quantum Dot with Fast rf Gate Sensors},
  author = {Colless, J. I. and Mahoney, A. C. and Hornibrook, J. M. and Doherty, A. C. and Lu, H. and Gossard, A. C. and Reilly, D. J.},
  journal = {Phys. Rev. Lett.},
  volume = {110},
  issue = {4},
  pages = {046805},
  numpages = {5},
  year = {2013},
  month = {Jan},
  publisher = {American Physical Society},
  doi = {10.1103/PhysRevLett.110.046805},
  url = {https://link.aps.org/doi/10.1103/PhysRevLett.110.046805}
}

@article{Spethmann,
  title = {Spin-qubit readout analysis based on a hidden {M}arkov model},
  author = {Spethmann, Maria and Stano, Peter and Loss, Daniel},
  journal = {Phys. Rev. B},
  volume = {112},
  issue = {11},
  pages = {115304},
  numpages = {14},
  year = {2025},
  month = {Sep},
  publisher = {American Physical Society},
  doi = {10.1103/x8yx-5111},
  url = {https://link.aps.org/doi/10.1103/x8yx-5111}
}

@article{KotekarPatilPSS2017,
   author = {D. Kotekar‐Patil and A. Corna and R. Maurand and A. Crippa and A. Orlov and S. Barraud and L. Hutin and M. Vinet and X. Jehl and S. De Franceschi and M. Sanquer},
   doi = {10.1002/pssb.201600581},
   issn = {0370-1972},
   issue = {3},
   journal = {physica status solidi (b)},
   month = {3},
   title = {Pauli spin blockade in {CMOS} double quantum dot devices},
   volume = {254},
   url = {https://onlinelibrary.wiley.com/doi/10.1002/pssb.201600581},
   year = {2017}
}

@article{MadzikNature2025,
   author = {Mateusz T. Mądzik and Florian Luthi and Gian Giacomo Guerreschi and Fahd A. Mohiyaddin and Felix Borjans and Jason D. Chadwick and Matthew J. Curry and Joshua Ziegler and Sarah Atanasov and Peter L. Bavdaz and Elliot J. Connors and J. Corrigan and H. Ekmel Ercan and Robert Flory and Hubert C. George and Benjamin Harpt and Eric Henry and Mohammad M. Islam and Nader Khammassi and Daniel Keith and Lester F. Lampert and Todor M. Mladenov and Randy W. Morris and Aditi Nethwewala and Samuel Neyens and René Otten and Linda P. Osuna Ibarra and Bishnu Patra and Ravi Pillarisetty and Shavindra Premaratne and Mick Ramsey and Andrew Risinger and John D. Rooney and Rostyslav Savytskyy and Thomas F. Watson and Otto K. Zietz and Anne Y. Matsuura and Stefano Pellerano and Nathaniel C. Bishop and Jeanette Roberts and James S. Clarke},
   doi = {10.1038/s41586-025-09767-5},
   issn = {0028-0836},
   issue = {8091},
   journal = {Nature},
   month = {11},
   pages = {870-875},
   title = {Operating two exchange-only qubits in parallel},
   volume = {647},
   year = {2025}
}

@article{George2025,
   author = {Hubert C. George and Mateusz T. Mądzik and Eric M. Henry and Andrew J. Wagner and Mohammad M. Islam and Felix Borjans and Elliot J. Connors and J. Corrigan and Matthew Curry and Michael K. Harper and Daniel Keith and Lester Lampert and Florian Luthi and Fahd A. Mohiyaddin and Sandra Murcia and Rohit Nair and Rambert Nahm and Aditi Nethwewala and Samuel Neyens and Bishnu Patra and Roy D. Raharjo and Carly Rogan and Rostyslav Savytskyy and Thomas F. Watson and Josh Ziegler and Otto K. Zietz and Stefano Pellerano and Ravi Pillarisetty and Nathaniel C. Bishop and Stephanie A. Bojarski and Jeanette Roberts and James S. Clarke},
   doi = {10.1021/acs.nanolett.4c05205},
   issn = {1530-6984},
   issue = {2},
   journal = {Nano Letters},
   month = {1},
   pages = {793-799},
   title = {12-Spin-Qubit Arrays Fabricated on a 300 mm Semiconductor Manufacturing Line},
   volume = {25},
   year = {2025}
}

@article{Neyens2024,
   author = {Samuel Neyens and Otto K. Zietz and Thomas F. Watson and Florian Luthi and Aditi Nethwewala and Hubert C. George and Eric Henry and Mohammad Islam and Andrew J. Wagner and Felix Borjans and Elliot J. Connors and J. Corrigan and Matthew J. Curry and Daniel Keith and Roza Kotlyar and Lester F. Lampert and Mateusz T. Mądzik and Kent Millard and Fahd A. Mohiyaddin and Stefano Pellerano and Ravi Pillarisetty and Mick Ramsey and Rostyslav Savytskyy and Simon Schaal and Guoji Zheng and Joshua Ziegler and Nathaniel C. Bishop and Stephanie Bojarski and Jeanette Roberts and James S. Clarke},
   doi = {10.1038/s41586-024-07275-6},
   issn = {0028-0836},
   issue = {8010},
   journal = {Nature},
   month = {5},
   pages = {80-85},
   title = {Probing single electrons across 300-mm spin qubit wafers},
   volume = {629},
   year = {2024}
}

@article{Zwerver2022,
   author = {A. M. J. Zwerver and T. Krähenmann and T. F. Watson and L. Lampert and H. C. George and R. Pillarisetty and S. A. Bojarski and P. Amin and S. V. Amitonov and J. M. Boter and R. Caudillo and D. Correas-Serrano and J. P. Dehollain and G. Droulers and E. M. Henry and R. Kotlyar and M. Lodari and F. Lüthi and D. J. Michalak and B. K. Mueller and S. Neyens and J. Roberts and N. Samkharadze and G. Zheng and O. K. Zietz and G. Scappucci and M. Veldhorst and L. M. K. Vandersypen and J. S. Clarke},
   doi = {10.1038/s41928-022-00727-9},
   issn = {2520-1131},
   issue = {3},
   journal = {Nature Electronics},
   month = {3},
   pages = {184-190},
   title = {Qubits made by advanced semiconductor manufacturing},
   volume = {5},
   year = {2022}
}

@article{ItohReview2014,
	author = {Itoh, Kohei M. and Watanabe, Hideyuki},
	journal = {MRS Communications},
	number = {4},
	pages = {143--157},
	title = {Isotope engineering of silicon and diamond for quantum computing and sensing applications},
    doi = {10.1557/mrc.2014.32},
	volume = {4},
	year = {2014}}

@article{Tyryshkin2012,
   author = {Alexei M. Tyryshkin and Shinichi Tojo and John J. L. Morton and Helge Riemann and Nikolai V. Abrosimov and Peter Becker and Hans-Joachim Pohl and Thomas Schenkel and Michael L. W. Thewalt and Kohei M. Itoh and S. A. Lyon},
   doi = {10.1038/nmat3182},
   issn = {1476-1122},
   issue = {2},
   journal = {Nature Materials},
   month = {2},
   pages = {143-147},
   title = {Electron spin coherence exceeding seconds in high-purity silicon},
   volume = {11},
   year = {2012}
}

@article{DehollainBell,
	author = {Dehollain, Juan P. and Simmons, Stephanie and Muhonen, Juha T. and Kalra, Rachpon and Laucht, Arne and Hudson, Fay and Itoh, Kohei M. and Jamieson, David N. and McCallum, Jeffrey C. and Dzurak, Andrew S. and Morello, Andrea},
    doi = {10.1038/nnano.2015.262},
	journal = {Nature Nanotechnology},
	number = {3},
	pages = {242--246},
	title = {Bell's inequality violation with spins in silicon},
	volume = {11},
	year = {2016}}

@article{Muhonen30sec,
	author = {Muhonen, Juha T. and Dehollain, Juan P. and Laucht, Arne and Hudson, Fay E. and Kalra, Rachpon and Sekiguchi, Takeharu and Itoh, Kohei M. and Jamieson, David N. and McCallum, Jeffrey C. and Dzurak, Andrew S. and Morello, Andrea},
    doi = {10.1038/nnano.2014.211},
	journal = {Nature Nanotechnology},
	number = {12},
	pages = {986--991},
	title = {Storing quantum information for 30 seconds in a nanoelectronic device},
	volume = {9},
	year = {2014}}

@article{VeldhorstTwoQubitLogic,
	author = {Veldhorst, M. and Yang, C. H. and Hwang, J. C. C. and Huang, W. and Dehollain, J. P. and Muhonen, J. T. and Simmons, S. and Laucht, A. and Hudson, F. E. and Itoh, K. M. and Morello, A. and Dzurak, A. S.},
    doi = {10.1038/nature15263},
	journal = {Nature},
	number = {7573},
	pages = {410--414},
	title = {A two-qubit logic gate in silicon},
	volume = {526},
	year = {2015}}

@article{VeldhorstAddressable,
	author = {Veldhorst, M. and Hwang, J. C. C. and Yang, C. H. and Leenstra, A. W. and de Ronde, B. and Dehollain, J. P. and Muhonen, J. T. and Hudson, F. E. and Itoh, K. M. and Morello, A. and Dzurak, A. S.},
    doi = {10.1038/nnano.2014.216},
	journal = {Nature Nanotechnology},
	number = {12},
	pages = {981--985},
	title = {An addressable quantum dot qubit with fault-tolerant control-fidelity},
	volume = {9},
	year = {2014}}

@misc{UndsethArxiv2026,
      title={Weight-four parity checks with silicon spin qubits}, 
      author={Brennan Undseth and Nicola Meggiato and Yi-Hsien Wu and Sam R. Katiraee-Far and Larysa Tryputen and Sander L. de Snoo and Davide Degli Esposti and Giordano Scappucci and Eliška Greplová and Lieven M. K. Vandersypen},
      year={2026},
      eprint={2601.23267},
      archivePrefix={arXiv},
      primaryClass={cond-mat.mes-hall},
      url={https://arxiv.org/abs/2601.23267}, 
}

@article{TakedaQEC2022,
   author = {Kenta Takeda and Akito Noiri and Takashi Nakajima and Takashi Kobayashi and Seigo Tarucha},
   doi = {10.1038/s41586-022-04986-6},
   issn = {0028-0836},
   issue = {7924},
   journal = {Nature},
   month = {8},
   pages = {682-686},
   title = {Quantum error correction with silicon spin qubits},
   volume = {608},
   year = {2022}
}

@article{vanRiggelenQEC2022,
   author = {F. van Riggelen and W. I. L. Lawrie and M. Russ and N. W. Hendrickx and A. Sammak and M. Rispler and B. M. Terhal and G. Scappucci and M. Veldhorst},
   doi = {10.1038/s41534-022-00639-8},
   issn = {2056-6387},
   issue = {1},
   journal = {npj Quantum Information},
   month = {10},
   pages = {124},
   title = {Phase flip code with semiconductor spin qubits},
   volume = {8},
   year = {2022}
}

@article{HendrickxNature2021,
   author = {Nico W. Hendrickx and William I. L. Lawrie and Maximilian Russ and Floor van Riggelen and Sander L. de Snoo and Raymond N. Schouten and Amir Sammak and Giordano Scappucci and Menno Veldhorst},
   doi = {10.1038/s41586-021-03332-6},
   issn = {0028-0836},
   issue = {7851},
   journal = {Nature},
   month = {3},
   pages = {580-585},
   title = {A four-qubit germanium quantum processor},
   volume = {591},
   url = {https://www.nature.com/articles/s41586-021-03332-6},
   year = {2021}
}

@article{XinZhang2025,
   author = {Xin Zhang and Elizaveta Morozova and Maximilian Rimbach-Russ and Daniel Jirovec and Tzu-Kan Hsiao and Pablo Cova Fariña and Chien-An Wang and Stefan D. Oosterhout and Amir Sammak and Giordano Scappucci and Menno Veldhorst and Lieven M. K. Vandersypen},
   doi = {10.1038/s41565-024-01817-9},
   issn = {1748-3387},
   issue = {2},
   journal = {Nature Nanotechnology},
   month = {2},
   pages = {209-215},
   title = {Universal control of four singlet–triplet qubits},
   volume = {20},
   year = {2025}
}

@article{HuangNature2019,
	author = {Huang, W. and Yang, C. H. and Chan, K. W. and Tanttu, T. and Hensen, B. and Leon, R. C. C. and Fogarty, M. A. and Hwang, J. C. C. and Hudson, F. E. and Itoh, K. M. and Morello, A. and Laucht, A. and Dzurak, A. S.},
    doi = {10.1038/s41586-019-1197-0},
	journal = {Nature},
	number = {7757},
	pages = {532--536},
	title = {Fidelity benchmarks for two-qubit gates in silicon},
	volume = {569},
	year = {2019}}

@misc{YiHsienWuArxiv2025,
      title={Simultaneous High-Fidelity Single-Qubit Gates in a Spin Qubit Array}, 
      author={Yi-Hsien Wu and Leon C. Camenzind and Patrick Bütler and Ik Kyeong Jin and Akito Noiri and Kenta Takeda and Takashi Nakajima and Takashi Kobayashi and Giordano Scappucci and Hsi-Sheng Goan and Seigo Tarucha},
      year={2025},
      eprint={2507.11918},
      archivePrefix={arXiv},
      primaryClass={quant-ph},
      url={https://arxiv.org/abs/2507.11918}, 
}

@article{KoppensNature2006,
   author = {F. H. L. Koppens and C. Buizert and K. J. Tielrooij and I. T. Vink and K. C. Nowack and T. Meunier and L. P. Kouwenhoven and L. M. K. Vandersypen},
   doi = {10.1038/nature05065},
   issn = {0028-0836},
   issue = {7104},
   journal = {Nature},
   month = {8},
   pages = {766-771},
   title = {Driven coherent oscillations of a single electron spin in a quantum dot},
   volume = {442},
   url = {https://www.nature.com/articles/nature05065},
   year = {2006}
}

@article{HansonReview,
  title = {Spins in few-electron quantum dots},
  author = {Hanson, R. and Kouwenhoven, L. P. and Petta, J. R. and Tarucha, S. and Vandersypen, L. M. K.},
  journal = {Rev. Mod. Phys.},
  volume = {79},
  issue = {4},
  pages = {1217--1265},
  numpages = {0},
  year = {2007},
  month = {Oct},
  publisher = {American Physical Society},
  doi = {10.1103/RevModPhys.79.1217},
  url = {https://link.aps.org/doi/10.1103/RevModPhys.79.1217}
}

@article{ChatterjeeReview,
   author = {Anasua Chatterjee and Paul Stevenson and Silvano De Franceschi and Andrea Morello and Nathalie P. de Leon and Ferdinand Kuemmeth},
   doi = {10.1038/s42254-021-00283-9},
   issn = {2522-5820},
   issue = {3},
   journal = {Nature Reviews Physics},
   month = {2},
   pages = {157-177},
   title = {Semiconductor qubits in practice},
   volume = {3},
   year = {2021}
}

@article{Loss1998,
   author = {Daniel Loss and David P. DiVincenzo},
   doi = {10.1103/PhysRevA.57.120},
   issn = {1050-2947},
   issue = {1},
   journal = {Physical Review A},
   month = {1},
   pages = {120-126},
   publisher = {American Physical Society},
   title = {Quantum computation with quantum dots},
   volume = {57},
   url = {https://link.aps.org/doi/10.1103/PhysRevA.57.120},
   year = {1998}
}

@article{OnoScience2002,
author = {K. Ono  and D. G. Austing  and Y. Tokura  and S. Tarucha },
title = {Current Rectification by Pauli Exclusion in a Weakly Coupled Double Quantum Dot System},
journal = {Science},
volume = {297},
number = {5585},
pages = {1313-1317},
year = {2002},
doi = {10.1126/science.1070958},
URL = {https://www.science.org/doi/abs/10.1126/science.1070958}
}

@article{Elzerman2004,
   author = {J. M. Elzerman and R. Hanson and L. H. Willems van Beveren and B. Witkamp and L. M. K. Vandersypen and L. P. Kouwenhoven},
   doi = {10.1038/nature02693},
   issn = {0028-0836},
   issue = {6998},
   journal = {Nature},
   month = {7},
   pages = {431-435},
   title = {Single-shot read-out of an individual electron spin in a quantum dot},
   volume = {430},
   year = {2004}
}

@article{Petta2005,
   author = {J R Petta and A C Johnson and J M Taylor and E A Laird and A Yacoby and M D Lukin and C M Marcus and M P Hanson and A C Gossard},
   doi = {10.1126/science.1116955},
   issue = {5744},
   journal = {Science},
   month = {9},
   pages = {2180-2184},
   publisher = {American Association for the Advancement of Science},
   title = {Coherent Manipulation of Coupled Electron Spins in Semiconductor Quantum Dots},
   volume = {309},
   url = {https://doi.org/10.1126/science.1116955},
   year = {2005}
}

@article{Jouravlev2006Transport,
  title = {Electron Transport in a Double Quantum Dot Governed by a Nuclear Magnetic Field},
  author = {Jouravlev, Oleg N. and Nazarov, Yuli V.},
  journal = {Phys. Rev. Lett.},
  volume = {96},
  issue = {17},
  pages = {176804},
  numpages = {4},
  year = {2006},
  month = {May},
  publisher = {American Physical Society},
  doi = {10.1103/PhysRevLett.96.176804},
  url = {https://link.aps.org/doi/10.1103/PhysRevLett.96.176804}
}

@article{Reilly2007RfQPC,
    author = {Reilly, D. J. and Marcus, C. M. and Hanson, M. P. and Gossard, A. C.},
    title = {Fast single-charge sensing with a rf quantum point contact},
    journal = {Applied Physics Letters},
    volume = {91},
    number = {16},
    pages = {162101},
    year = {2007},
    month = {10},
    issn = {0003-6951},
    doi = {10.1063/1.2794995},
    url = {https://doi.org/10.1063/1.2794995}
}

@article{Danon2009psbStrongSOC,
   author = {J. Danon and Yu. V. Nazarov},
   doi = {10.1103/PhysRevB.80.041301},
   issn = {1098-0121},
   issue = {4},
   journal = {Physical Review B},
   month = {7},
   pages = {041301},
   publisher = {American Physical Society},
   title = {Pauli spin blockade in the presence of strong spin-orbit coupling},
   volume = {80},
   url = {https://link.aps.org/doi/10.1103/PhysRevB.80.041301},
   year = {2009}
}

@article{Barthel2009SingleShot,
  title = {Rapid Single-Shot Measurement of a Singlet-Triplet Qubit},
  author = {Barthel, C. and Reilly, D. J. and Marcus, C. M. and Hanson, M. P. and Gossard, A. C.},
  journal = {Phys. Rev. Lett.},
  volume = {103},
  issue = {16},
  pages = {160503},
  numpages = {4},
  year = {2009},
  month = {Oct},
  publisher = {American Physical Society},
  doi = {10.1103/PhysRevLett.103.160503},
  url = {https://link.aps.org/doi/10.1103/PhysRevLett.103.160503}
}

@article{Barthel2010,
   author = {C. Barthel and M. Kjærgaard and J. Medford and M. Stopa and C. M. Marcus and M. P. Hanson and A. C. Gossard},
   doi = {10.1103/PhysRevB.81.161308},
   issn = {1098-0121},
   issue = {16},
   journal = {Physical Review B},
   month = {4},
   pages = {161308},
   title = {Fast sensing of double-dot charge arrangement and spin state with a radio-frequency sensor quantum dot},
   volume = {81},
   year = {2010}
}

@article{Zwanenburg2013SiQuantElec,
   author = {Floris A. Zwanenburg and Andrew S. Dzurak and Andrea Morello and Michelle Y. Simmons and Lloyd C. L. Hollenberg and Gerhard Klimeck and Sven Rogge and Susan N. Coppersmith and Mark A. Eriksson},
   doi = {10.1103/RevModPhys.85.961},
   issn = {0034-6861},
   issue = {3},
   journal = {Reviews of Modern Physics},
   month = {7},
   pages = {961-1019},
   publisher = {American Physical Society},
   title = {Silicon quantum electronics},
   volume = {85},
   url = {https://link.aps.org/doi/10.1103/RevModPhys.85.961},
   year = {2013}
}

@article{Tahan2014Relaxation,
  title = {Relaxation of excited spin, orbital, and valley qubit states in ideal silicon quantum dots},
  author = {Tahan, Charles and Joynt, Robert},
  journal = {Phys. Rev. B},
  volume = {89},
  issue = {7},
  pages = {075302},
  numpages = {22},
  year = {2014},
  month = {Feb},
  publisher = {American Physical Society},
  doi = {10.1103/PhysRevB.89.075302},
  url = {https://link.aps.org/doi/10.1103/PhysRevB.89.075302}
}

@article{Maurand2016cmos,
   author = {R. Maurand and X. Jehl and D. Kotekar-Patil and A. Corna and H. Bohuslavskyi and R. Laviéville and L. Hutin and S. Barraud and M. Vinet and M. Sanquer and S. De Franceschi},
   doi = {10.1038/ncomms13575},
   issn = {2041-1723},
   issue = {1},
   journal = {Nature Communications},
   month = {11},
   pages = {13575},
   title = {A {CMOS} silicon spin qubit},
   volume = {7},
   year = {2016}
}

@article{Boross2016Nanotech,
    doi = {10.1088/0957-4484/27/31/314002},
    url = {https://doi.org/10.1088/0957-4484/27/31/314002},
    year = {2016},
    month = {jun},
    publisher = {IOP Publishing},
    volume = {27},
    number = {31},
    pages = {314002},
    author = {Boross, Péter and Széchenyi, Gábor and Pályi, András},
    title = {Valley-enhanced fast relaxation of gate-controlled donor qubits in silicon},
    journal = {Nanotechnology},
}

@article{Boross2016ValleySi,
  title = {Control of valley dynamics in silicon quantum dots in the presence of an interface step},
  author = {Boross, P\'eter and Sz\'echenyi, G\'abor and Culcer, Dimitrie and P\'alyi, Andr\'as},
  journal = {Phys. Rev. B},
  volume = {94},
  issue = {3},
  pages = {035438},
  numpages = {11},
  year = {2016},
  month = {Jul},
  publisher = {American Physical Society},
  doi = {10.1103/PhysRevB.94.035438},
  url = {https://link.aps.org/doi/10.1103/PhysRevB.94.035438}
}

@article{Mizuta2017,
   author = {R. Mizuta and R. M. Otxoa and A. C. Betz and M. F. Gonzalez-Zalba},
   doi = {10.1103/PhysRevB.95.045414},
   issn = {2469-9950},
   issue = {4},
   journal = {Physical Review B},
   month = {1},
   pages = {045414},
   title = {Quantum and tunneling capacitance in charge and spin qubits},
   volume = {95},
   year = {2017}
}

@article{vandersypen2017Scalling,
   author = {L. M. K. Vandersypen and H. Bluhm and J. S. Clarke and A. S. Dzurak and R. Ishihara and A. Morello and D. J. Reilly and L. R. Schreiber and M. Veldhorst},
   doi = {10.1038/s41534-017-0038-y},
   issn = {2056-6387},
   issue = {1},
   journal = {npj Quantum Information},
   month = {9},
   pages = {34},
   title = {Interfacing spin qubits in quantum dots and donors—hot, dense, and coherent},
   volume = {3},
   year = {2017}
}

@article{Yoneda2018,
   author = {Jun Yoneda and Kenta Takeda and Tomohiro Otsuka and Takashi Nakajima and Matthieu R. Delbecq and Giles Allison and Takumu Honda and Tetsuo Kodera and Shunri Oda and Yusuke Hoshi and Noritaka Usami and Kohei M. Itoh and Seigo Tarucha},
   doi = {10.1038/s41565-017-0014-x},
   issn = {1748-3387},
   issue = {2},
   journal = {Nature Nanotechnology},
   month = {2},
   pages = {102-106},
   title = {A quantum-dot spin qubit with coherence limited by charge noise and fidelity higher than 99.9\%},
   volume = {13},
   year = {2018}
}

@article{Kornich2018RelaxationSiGe,
  doi = {10.22331/q-2018-05-28-70},
  url = {https://doi.org/10.22331/q-2018-05-28-70},
  title = {Phonon-assisted relaxation and decoherence of singlet-triplet qubits in {S}i/{S}i{G}e quantum dots},
  author = {Kornich, Viktoriia and Kloeffel, Christoph and Loss, Daniel},
  journal = {{Quantum}},
  issn = {2521-327X},
  publisher = {{Verein zur F{\"{o}}rderung des Open Access Publizierens in den Quantenwissenschaften}},
  volume = {2},
  pages = {70},
  month = may,
  year = {2018}
}

@article{West2019SingleShotRF,
   author = {Anderson West and Bas Hensen and Alexis Jouan and Tuomo Tanttu and Chih-Hwan Yang and Alessandro Rossi and M. Fernando Gonzalez-Zalba and Fay Hudson and Andrea Morello and David J. Reilly and Andrew S. Dzurak},
   doi = {10.1038/s41565-019-0400-7},
   issn = {1748-3387},
   issue = {5},
   journal = {Nature Nanotechnology},
   month = {5},
   pages = {437-441},
   title = {Gate-based single-shot readout of spins in silicon},
   volume = {14},
   year = {2019}
}

@article{Esterli2019,
   author = {M. Esterli and R. M. Otxoa and M. F. Gonzalez-Zalba},
   doi = {10.1063/1.5098889},
   issn = {0003-6951},
   issue = {25},
   journal = {Applied Physics Letters},
   month = {6},
   pages = {253505},
   title = {Small-signal equivalent circuit for double quantum dots at low-frequencies},
   volume = {114},
   year = {2019}
}

@article{Crippa2019reflectometrySi,
   author = {A. Crippa and R. Ezzouch and A. Aprá and A. Amisse and R. Laviéville and L. Hutin and B. Bertrand and M. Vinet and M. Urdampilleta and T. Meunier and M. Sanquer and X. Jehl and R. Maurand and S. De Franceschi},
   doi = {10.1038/s41467-019-10848-z},
   issn = {2041-1723},
   issue = {1},
   journal = {Nature Communications},
   month = {7},
   pages = {2776},
   title = {Gate-reflectometry dispersive readout and coherent control of a spin qubit in silicon},
   volume = {10},
   year = {2019}
}

@article{Urdampilleta2019CMOSGateRf,
   author = {Matias Urdampilleta and David J. Niegemann and Emmanuel Chanrion and Baptiste Jadot and Cameron Spence and Pierre-André Mortemousque and Christopher Bäuerle and Louis Hutin and Benoit Bertrand and Sylvain Barraud and Romain Maurand and Marc Sanquer and Xavier Jehl and Silvano De Franceschi and Maud Vinet and Tristan Meunier},
   doi = {10.1038/s41565-019-0443-9},
   issn = {1748-3387},
   issue = {8},
   journal = {Nature Nanotechnology},
   month = {8},
   pages = {737-741},
   title = {Gate-based high fidelity spin readout in a {CMOS} device},
   volume = {14},
   year = {2019}
}

@article{Zheng2019rapidreadout,
   author = {Guoji Zheng and Nodar Samkharadze and Marc L. Noordam and Nima Kalhor and Delphine Brousse and Amir Sammak and Giordano Scappucci and Lieven M. K. Vandersypen},
   doi = {10.1038/s41565-019-0488-9},
   issn = {1748-3387},
   issue = {8},
   journal = {Nature Nanotechnology},
   month = {8},
   pages = {742-746},
   title = {Rapid gate-based spin read-out in silicon using an on-chip resonator},
   volume = {14},
   year = {2019}
}

@article{DanjouPRBDispersiveReadout,
  title = {Optimal dispersive readout of a spin qubit with a microwave resonator},
  author = {D'Anjou, B. and Burkard, Guido},
  journal = {Phys. Rev. B},
  volume = {100},
  issue = {24},
  pages = {245427},
  numpages = {17},
  year = {2019},
  month = {Dec},
  publisher = {American Physical Society},
  doi = {10.1103/PhysRevB.100.245427},
  url = {https://link.aps.org/doi/10.1103/PhysRevB.100.245427}
}

@article{AhmedRFSensing,
  title = {Radio-Frequency Capacitive Gate-Based Sensing},
  author = {Ahmed, Imtiaz and Haigh, James A. and Schaal, Simon and Barraud, Sylvain and Zhu, Yi and Lee, Chang-min and Amado, Mario and Robinson, Jason W. A. and Rossi, Alessandro and Morton, John J. L. and Gonzalez-Zalba, M. Fernando},
  journal = {Phys. Rev. Appl.},
  volume = {10},
  issue = {1},
  pages = {014018},
  numpages = {9},
  year = {2018},
  month = {Jul},
  publisher = {American Physical Society},
  doi = {10.1103/PhysRevApplied.10.014018},
  url = {https://link.aps.org/doi/10.1103/PhysRevApplied.10.014018}
}

@article{Vahid2020,
   author = {Vahid Derakhshan Maman and M.F. Gonzalez-Zalba and András Pályi},
   doi = {10.1103/PhysRevApplied.14.064024},
   issn = {2331-7019},
   issue = {6},
   journal = {Physical Review Applied},
   month = {12},
   pages = {064024},
   title = {Charge Noise and Overdrive Errors in Dispersive Readout of Charge, Spin, and {M}ajorana Qubits},
   volume = {14},
   year = {2020}
}

@article{Yoneda2020QNDSi,
   author = {J. Yoneda and K. Takeda and A. Noiri and T. Nakajima and S. Li and J. Kamioka and T. Kodera and S. Tarucha},
   doi = {10.1038/s41467-020-14818-8},
   issn = {2041-1723},
   issue = {1},
   journal = {Nature Communications},
   month = {3},
   pages = {1144},
   title = {Quantum non-demolition readout of an electron spin in silicon},
   volume = {11},
   year = {2020}
}

@article{Connors2020GateRfRapid,
  title = {Rapid High-Fidelity Spin-State Readout in $\mathrm{Si}$/$\mathrm{Si}$-$\mathrm{Ge}$ Quantum Dots via rf Reflectometry},
  author = {Connors, Elliot J. and Nelson, JJ and Nichol, John M.},
  journal = {Phys. Rev. Appl.},
  volume = {13},
  issue = {2},
  pages = {024019},
  numpages = {9},
  year = {2020},
  month = {Feb},
  publisher = {American Physical Society},
  doi = {10.1103/PhysRevApplied.13.024019},
  url = {https://link.aps.org/doi/10.1103/PhysRevApplied.13.024019}
}

@article{Jang2020RfSET,
   author = {Wonjin Jang and Jehyun Kim and Min-Kyun Cho and Hwanchul Chung and Sanghyeok Park and Jaeun Eom and Vladimir Umansky and Yunchul Chung and Dohun Kim},
   doi = {10.1038/s41534-020-00295-w},
   issn = {2056-6387},
   issue = {1},
   journal = {npj Quantum Information},
   month = {7},
   pages = {64},
   title = {Robust energy-selective tunneling readout of singlet-triplet qubits under large magnetic field gradient},
   volume = {6},
   year = {2020}
}

@article{Yoneda2021,
   author = {J. Yoneda and W. Huang and M. Feng and C. H. Yang and K. W. Chan and T. Tanttu and W. Gilbert and R. C. C. Leon and F. E. Hudson and K. M. Itoh and A. Morello and S. D. Bartlett and A. Laucht and A. Saraiva and A. S. Dzurak},
   doi = {10.1038/s41467-021-24371-7},
   issn = {2041-1723},
   issue = {1},
   journal = {Nature Communications},
   month = {7},
   pages = {4114},
   title = {Coherent spin qubit transport in silicon},
   volume = {12},
   url = {https://www.nature.com/articles/s41467-021-24371-7},
   year = {2021}
}

@article{Ciriano-Tejel2021,
   author = {Virginia N. Ciriano-Tejel and Michael A. Fogarty and Simon Schaal and Louis Hutin and Benoit Bertrand and Lisa Ibberson and M. Fernando Gonzalez-Zalba and Jing Li and Yann-Michel Niquet and Maud Vinet and John J.L. Morton},
   doi = {10.1103/PRXQuantum.2.010353},
   issn = {2691-3399},
   issue = {1},
   journal = {PRX Quantum},
   month = {3},
   pages = {010353},
   title = {Spin Readout of a {CMOS} Quantum Dot by Gate Reflectometry and Spin-Dependent Tunneling},
   volume = {2},
   year = {2021}
}

@article{Liu2021,
   title = {Magnetic-Gradient-Free Two-Axis Control of a Valley Spin Qubit in $\mathrm{Si}_x\mathrm{Ge}_{1-x}$},
   author = {Y.-Y. Liu and L.A. Orona and Samuel F. Neyens and E.R. MacQuarrie and M.A. Eriksson and A. Yacoby},
   doi = {10.1103/PhysRevApplied.16.024029},
   issn = {2331-7019},
   issue = {2},
   journal = {Physical Review Applied},
   month = {8},
   pages = {024029},
   volume = {16},
   year = {2021}
}

@article{Philips2022,
   author = {Stephan G J Philips and Mateusz T Mądzik and Sergey V Amitonov and Sander L de Snoo and Maximilian Russ and Nima Kalhor and Christian Volk and William I L Lawrie and Delphine Brousse and Larysa Tryputen and Brian Paquelet Wuetz and Amir Sammak and Menno Veldhorst and Giordano Scappucci and Lieven M K Vandersypen},
   doi = {10.1038/s41586-022-05117-x},
   issn = {1476-4687},
   issue = {7929},
   journal = {Nature},
   pages = {919-924},
   title = {Universal control of a six-qubit quantum processor in silicon},
   volume = {609},
   url = {https://doi.org/10.1038/s41586-022-05117-x},
   year = {2022}
}

@article{Stano2022,
   author = {Peter Stano and Daniel Loss},
   doi = {10.1038/s42254-022-00484-w},
   issn = {2522-5820},
   issue = {10},
   journal = {Nature Reviews Physics},
   month = {8},
   pages = {672-688},
   title = {Review of performance metrics of spin qubits in gated semiconducting nanostructures},
   volume = {4},
   year = {2022}
}

@article{Connors2022,
   author = {Elliot J. Connors and J. Nelson and Lisa F. Edge and John M. Nichol},
   doi = {10.1038/s41467-022-28519-x},
   issn = {2041-1723},
   issue = {1},
   journal = {Nature Communications},
   month = {2},
   pages = {940},
   title = {Charge-noise spectroscopy of $\mathrm{Si}/\mathrm{SiGe}$ quantum dots via dynamically-decoupled exchange oscillations},
   volume = {13},
   year = {2022}
}

@article{Boter2022,
   author = {Jelmer M. Boter and Juan P. Dehollain and Jeroen P.G. van Dijk and Yuanxing Xu and Toivo Hensgens and Richard Versluis and Henricus W.L. Naus and James S. Clarke and Menno Veldhorst and Fabio Sebastiano and Lieven M.K. Vandersypen},
   doi = {10.1103/PhysRevApplied.18.024053},
   issn = {2331-7019},
   issue = {2},
   journal = {Physical Review Applied},
   month = {8},
   pages = {024053},
   title = {Spiderweb Array: A Sparse Spin-Qubit Array},
   volume = {18},
   year = {2022}
}

@article{Xue2022,
   author = {Xiao Xue and Maximilian Russ and Nodar Samkharadze and Brennan Undseth and Amir Sammak and Giordano Scappucci and Lieven M. K. Vandersypen},
   doi = {10.1038/s41586-021-04273-w},
   issn = {0028-0836},
   issue = {7893},
   journal = {Nature},
   month = {1},
   pages = {343-347},
   title = {Quantum logic with spin qubits crossing the surface code threshold},
   volume = {601},
   year = {2022}
}

@article{Mills2022,
   author = {A.R. Mills and C.R. Guinn and M.M. Feldman and A.J. Sigillito and M.J. Gullans and M.T. Rakher and J. Kerckhoff and C.A.C. Jackson and J.R. Petta},
   doi = {10.1103/PhysRevApplied.18.064028},
   issn = {2331-7019},
   issue = {6},
   journal = {Physical Review Applied},
   month = {12},
   pages = {064028},
   title = {High-Fidelity State Preparation, Quantum Control, and Readout of an Isotopically Enriched Silicon Spin Qubit},
   volume = {18},
   year = {2022}
}

@article{Fehse2023StateTransfer,
  title = {Generalized fast quasiadiabatic population transfer for improved qubit readout, shuttling, and noise mitigation},
  author = {Fehse, F. and David, M. and Pioro-Ladri\`ere, M. and Coish, W. A.},
  journal = {Phys. Rev. B},
  volume = {107},
  issue = {24},
  pages = {245303},
  numpages = {14},
  year = {2023},
  month = {Jun},
  publisher = {American Physical Society},
  doi = {10.1103/PhysRevB.107.245303},
  url = {https://link.aps.org/doi/10.1103/PhysRevB.107.245303}
}

@article{Nurizzo2023,
   author = {Martin Nurizzo and Baptiste Jadot and Pierre-André Mortemousque and Vivien Thiney and Emmanuel Chanrion and David Niegemann and Matthieu Dartiailh and Arne Ludwig and Andreas D. Wieck and Christopher Bäuerle and Matias Urdampilleta and Tristan Meunier},
   doi = {10.1103/PRXQuantum.4.010329},
   issn = {2691-3399},
   issue = {1},
   journal = {PRX Quantum},
   month = {3},
   pages = {010329},
   title = {Complete Readout of Two-Electron Spin States in a Double Quantum Dot},
   volume = {4},
   year = {2023}
}

@article{Burkard2023review,
   author = {Guido Burkard and Thaddeus D. Ladd and Andrew Pan and John M. Nichol and Jason R. Petta},
   doi = {10.1103/RevModPhys.95.025003},
   issn = {0034-6861},
   issue = {2},
   journal = {Reviews of Modern Physics},
   month = {6},
   pages = {025003},
   title = {Semiconductor spin qubits},
   volume = {95},
   year = {2023}
}

@article{Vigneau2023,
    author = {Vigneau, Florian and Fedele, Federico and Chatterjee, Anasua and Reilly, David and Kuemmeth, Ferdinand and Gonzalez-Zalba, M. Fernando and Laird, Edward and Ares, Natalia},
    title = {Probing quantum devices with radio-frequency reflectometry},
    journal = {Applied Physics Reviews},
    volume = {10},
    number = {2},
    pages = {021305},
    year = {2023},
    month = {02},
    issn = {1931-9401},
    doi = {10.1063/5.0088229},
    url = {https://doi.org/10.1063/5.0088229}
}

@article{Sen2023,
   author = {Aritra Sen and György Frank and Baksa Kolok and Jeroen Danon and András Pályi},
   doi = {10.1103/PhysRevB.108.245406},
   issn = {2469-9950},
   issue = {24},
   journal = {Physical Review B},
   month = {12},
   pages = {245406},
   title = {Classification and magic magnetic field directions for spin-orbit-coupled double quantum dots},
   volume = {108},
   url = {https://link.aps.org/doi/10.1103/PhysRevB.108.245406},
   year = {2023}
}

@article{Russell2023GateRfSiHole,
  title = {Gate-Based Spin Readout of Hole Quantum Dots with Site-Dependent $g$-Factors},
  author = {Russell, Angus and Zotov, Alexander and Zhao, Ruichen and Dzurak, Andrew S. and Fernando Gonzalez-Zalba, M. and Rossi, Alessandro},
  journal = {Phys. Rev. Appl.},
  volume = {19},
  issue = {4},
  pages = {044039},
  numpages = {11},
  year = {2023},
  month = {Apr},
  publisher = {American Physical Society},
  doi = {10.1103/PhysRevApplied.19.044039},
  url = {https://link.aps.org/doi/10.1103/PhysRevApplied.19.044039}
}

@article{Huang2024Nature1KSiMOS,
   author = {Jonathan Y. Huang and Rocky Y. Su and Wee Han Lim and MengKe Feng and Barnaby van Straaten and Brandon Severin and Will Gilbert and Nard Dumoulin Stuyck and Tuomo Tanttu and Santiago Serrano and Jesus D. Cifuentes and Ingvild Hansen and Amanda E. Seedhouse and Ensar Vahapoglu and Ross C. C. Leon and Nikolay V. Abrosimov and Hans-Joachim Pohl and Michael L. W. Thewalt and Fay E. Hudson and Christopher C. Escott and Natalia Ares and Stephen D. Bartlett and Andrea Morello and Andre Saraiva and Arne Laucht and Andrew S. Dzurak and Chih Hwan Yang},
   doi = {10.1038/s41586-024-07160-2},
   issn = {0028-0836},
   issue = {8005},
   journal = {Nature},
   month = {3},
   pages = {772-777},
   title = {High-fidelity spin qubit operation and algorithmic initialization above 1 {K}},
   volume = {627},
   year = {2024}
}

@article{Patomki2024SiPipeline,
   author = {S. M. Patomäki and M. F. Gonzalez-Zalba and M. A. Fogarty and Z. Cai and S. C. Benjamin and J. J. L. Morton},
   doi = {10.1038/s41534-024-00823-y},
   issn = {2056-6387},
   issue = {1},
   journal = {npj Quantum Information},
   month = {3},
   pages = {31},
   title = {Pipeline quantum processor architecture for silicon spin qubits},
   volume = {10},
   year = {2024}
}

@article{Kunne2024SpinBus,
   author = {Matthias Künne and Alexander Willmes and Max Oberländer and Christian Gorjaew and Julian D. Teske and Harsh Bhardwaj and Max Beer and Eugen Kammerloher and René Otten and Inga Seidler and Ran Xue and Lars R. Schreiber and Hendrik Bluhm},
   doi = {10.1038/s41467-024-49182-4},
   issn = {2041-1723},
   issue = {1},
   journal = {Nature Communications},
   month = {6},
   pages = {4977},
   title = {The {SpinBus} architecture for scaling spin qubits with electron shuttling},
   volume = {15},
   year = {2024}
}

@article{Peri2024Unified,
   author = {L. Peri and M. Benito and C. J. B. Ford and M. F. Gonzalez-Zalba},
   doi = {10.1038/s41534-024-00907-9},
   issn = {2056-6387},
   issue = {1},
   journal = {npj Quantum Information},
   month = {11},
   pages = {114},
   title = {Unified linear response theory of quantum electronic circuits},
   volume = {10},
   year = {2024}
}

@article{Lundberg2024LiftedPSB,
   author = {Theodor Lundberg and David J. Ibberson and Jing Li and Louis Hutin and José C. Abadillo-Uriel and Michele Filippone and Benoit Bertrand and Andreas Nunnenkamp and Chang-Min Lee and Nadia Stelmashenko and Jason W. A. Robinson and Maud Vinet and Lisa Ibberson and Yann-Michel Niquet and M. Fernando Gonzalez-Zalba},
   doi = {10.1038/s41534-024-00820-1},
   issn = {2056-6387},
   issue = {1},
   journal = {npj Quantum Information},
   month = {3},
   pages = {28},
   title = {Non-symmetric {Pauli} spin blockade in a silicon double quantum dot},
   volume = {10},
   year = {2024}
}

@article{Unseld2025,
   author = {Florian K. Unseld and Brennan Undseth and Eline Raymenants and Yuta Matsumoto and Sander L. de Snoo and Saurabh Karwal and Oriol Pietx-Casas and Alexander S. Ivlev and Marcel Meyer and Amir Sammak and Menno Veldhorst and Giordano Scappucci and Lieven M. K. Vandersypen},
   doi = {10.1038/s41467-025-60351-x},
   issn = {2041-1723},
   issue = {1},
   journal = {Nature Communications},
   month = {7},
   pages = {5605},
   title = {Baseband control of single-electron silicon spin qubits in two dimensions},
   volume = {16},
   year = {2025}
}

@article{Steinacker2025,
   author = {Paul Steinacker and Nard Dumoulin Stuyck and Wee Han Lim and Tuomo Tanttu and MengKe Feng and Santiago Serrano and Andreas Nickl and Marco Candido and Jesus D. Cifuentes and Ensar Vahapoglu and Samuel K. Bartee and Fay E. Hudson and Kok Wai Chan and Stefan Kubicek and Julien Jussot and Yann Canvel and Sofie Beyne and Yosuke Shimura and Roger Loo and Clement Godfrin and Bart Raes and Sylvain Baudot and Danny Wan and Arne Laucht and Chih Hwan Yang and Andre Saraiva and Christopher C. Escott and Kristiaan De Greve and Andrew S. Dzurak},
   doi = {10.1038/s41586-025-09531-9},
   issn = {0028-0836},
   issue = {8083},
   journal = {Nature},
   month = {10},
   pages = {81-87},
   title = {Industry-compatible silicon spin-qubit unit cells exceeding 99\% fidelity},
   volume = {646},
   year = {2025}
}

@article{Bartee2025,
   author = {Samuel K. Bartee and Will Gilbert and Kun Zuo and Kushal Das and Tuomo Tanttu and Chih Hwan Yang and Nard Dumoulin Stuyck and Sebastian J. Pauka and Rocky Y. Su and Wee Han Lim and Santiago Serrano and Christopher C. Escott and Fay E. Hudson and Kohei M. Itoh and Arne Laucht and Andrew S. Dzurak and David J. Reilly},
   doi = {10.1038/s41586-025-09157-x},
   issn = {0028-0836},
   issue = {8071},
   journal = {Nature},
   month = {7},
   pages = {382-387},
   title = {Spin-qubit control with a milli-kelvin {CMOS} chip},
   volume = {643},
   year = {2025}
}

@article{Pataki2025,
   author = {Dávid Pataki and András Pályi},
   doi = {10.1103/PhysRevB.111.115307},
   issn = {2469-9950},
   issue = {11},
   journal = {Physical Review B},
   month = {3},
   pages = {115307},
   title = {Compiling the surface code to crossbar spin qubit architectures},
   volume = {111},
   year = {2025}
}

@article{Peri2025MicrowaveSim,
   author = {Lorenzo Peri and Alberto Gomez-Saiz and Christopher J. B. Ford and M. Fernando Gonzalez-Zalba},
   doi = {10.1038/s41534-025-01140-8},
   issn = {2056-6387},
   issue = {1},
   journal = {npj Quantum Information},
   month = {12},
   pages = {194},
   title = {Compact quantum dot models for analog microwave co-simulation},
   volume = {11},
   year = {2025}
}

@article{vonHorstig2025,
   author = {Felix-Ekkehard von Horstig and Lorenzo Peri and Virginia N. Ciriano-Tejel and Sylvain Barraud and Jason A. W. Robinson and Monica Benito and Frederico Martins and M. Fernando Gonzalez-Zalba},
   doi = {10.1038/s41534-025-01102-0},
   issn = {2056-6387},
   issue = {1},
   journal = {npj Quantum Information},
   month = {10},
   pages = {155},
   title = {Electrical readout of spins in the absence of spin blockade},
   volume = {11},
   year = {2025}
}

@article{Ventura-Meinersen2025Optimization,
   author = {Chris Ventura-Meinersen and Stefano Bosco and Maximilian Rimbach-Russ},
   doi = {10.1140/epjqt/s40507-025-00426-2},
   issn = {2662-4400},
   issue = {1},
   journal = {EPJ Quantum Technology},
   month = {12},
   pages = {125},
   title = {Quantum geometric protocols for fast high-fidelity adiabatic state transfer},
   volume = {12},
   year = {2025}
}

@misc{katiraeeFar2025Optimization,
      title={Unified evolutionary optimization for high-fidelity spin qubit operations}, 
      author={Sam R. Katiraee-Far and Yuta Matsumoto and Brennan Undseth and Maxim De Smet and Valentina Gualtieri and Christian Ventura Meinersen and Irene Fernandez de Fuentes and Kenji Capannelli and Maximilian Rimbach-Russ and Giordano Scappucci and Lieven M. K. Vandersypen and Eliska Greplova},
      year={2025},
      eprint={2503.12256},
      archivePrefix={arXiv},
      primaryClass={quant-ph},
      url={https://arxiv.org/abs/2503.12256}, 
}

@article{FernndezdeFuentes2026,
   author = {I. Fernández de Fuentes and E. Raymenants and B. Undseth and O. Pietx-Casas and S. Philips and M. Mądzik and S.L. de Snoo and S.V. Amitonov and L. Tryputen and A.T. Schmitz and A.Y. Matsuura and G. Scappucci and L.M.K. Vandersypen},
   doi = {10.1103/f285-l2v5},
   issn = {2691-3399},
   issue = {1},
   journal = {PRX Quantum},
   month = {1},
   pages = {010308},
   title = {Running a Six-Qubit Quantum Circuit on a Silicon Spin-Qubit Array},
   volume = {7},
   year = {2026}
}

@misc{Laine2025dispersiveHMM,
      title={High-fidelity dispersive spin sensing in a tuneable unit cell of silicon {MOS} quantum dots}, 
      author={Constance Lainé and Giovanni A. Oakes and Virginia Ciriano-Tejel and Jacob F. Chittock-Wood and Lorenzo Peri and Michael A. Fogarty and Sofia M. Patomäki and Stefan Kubicek and David F. Wise and Ross C. C. Leon and M. Fernando Gonzalez-Zalba and John J. L. Morton},
      year={2025},
      eprint={2505.10435},
      archivePrefix={arXiv},
      primaryClass={quant-ph},
      url={https://arxiv.org/abs/2505.10435}, 
}

@misc{unseld2025Zenodo,
  author       = {Unseld, Florian Karl and
                  Undseth, Brennan and
                  Raymenants, Eline and
                  Matsumoto, Yuta and
                  Karwal, Saurabh and
                  Pietx-Casas, Oriol and
                  Ivlev, Alexander S. and
                  Meyer, Marcel and
                  Sammak, Amir and
                  Veldhorst, Menno and
                  Scappucci, Giordano and
                  Vandersypen, Lieven M. K.},
  title        = {Baseband control of single-electron silicon spin
                   qubits in two dimensions
                  },
  month        = dec,
  year         = 2024,
  publisher    = {Zenodo},
  doi          = {10.5281/zenodo.14234481},
  url          = {https://doi.org/10.5281/zenodo.14234481},
}

@article{Herring1957,
  title = {Transport and Deformation-Potential Theory for Many-Valley Semiconductors with Anisotropic Scattering},
  author = {Herring, Conyers and Vogt, Erich},
  journal = {Phys. Rev.},
  volume = {101},
  issue = {3},
  pages = {944--961},
  numpages = {0},
  year = {1956},
  month = {Feb},
  publisher = {American Physical Society},
  doi = {10.1103/PhysRev.101.944},
  url = {https://link.aps.org/doi/10.1103/PhysRev.101.944}
}

\end{document}